\documentclass[twoside,twocolumn,9pt]{article}
\usepackage{extsizes}
\usepackage[super,sort&compress,comma]{natbib} 
\usepackage[version=3]{mhchem}
\usepackage[left=1.5cm, right=1.5cm, top=1.785cm, bottom=2.0cm]{geometry}
\usepackage{balance}
\usepackage{mathptmx}
\usepackage{sectsty}
\usepackage{graphicx} 
\usepackage{lastpage}
\usepackage[format=plain,justification=justified,singlelinecheck=false,font={stretch=1.125,small,sf},labelfont=bf,labelsep=space]{caption}
\usepackage{float}
\usepackage{fancyhdr}
\usepackage{fnpos}
\usepackage[english]{babel}
\addto{\captionsenglish}{%
  
}
\usepackage{array}
\usepackage{droidsans}
\usepackage{charter}
\usepackage[T1]{fontenc}
\usepackage[usenames,dvipsnames]{xcolor}
\usepackage{setspace}
\usepackage[compact]{titlesec}
\usepackage{hyperref}
\usepackage{graphicx,amsmath,amssymb,color,soul}

\usepackage{epstopdf}

\definecolor{cream}{RGB}{222,217,201}

\begin{document}

\pagestyle{fancy}
\thispagestyle{plain}
\fancypagestyle{plain}{
\renewcommand{\headrulewidth}{0pt}
}

\makeFNbottom
\makeatletter
\renewcommand\LARGE{\@setfontsize\LARGE{15pt}{17}}
\renewcommand\Large{\@setfontsize\Large{12pt}{14}}
\renewcommand\large{\@setfontsize\large{10pt}{12}}
\renewcommand\footnotesize{\@setfontsize\footnotesize{7pt}{10}}
\makeatother

\renewcommand{\thefootnote}{\fnsymbol{footnote}}
\renewcommand\footnoterule{\vspace*{1pt}%
\color{cream}\hrule width 3.5in height 0.4pt \color{black}\vspace*{5pt}} 
\setcounter{secnumdepth}{5}

\makeatletter 
\renewcommand\@biblabel[1]{#1}            
\renewcommand\@makefntext[1]%
{\noindent\makebox[0pt][r]{\@thefnmark\,}#1}
\makeatother 
\renewcommand{\figurename}{\small{Fig.}~}
\sectionfont{\sffamily\Large}
\subsectionfont{\normalsize}
\subsubsectionfont{\bf}
\setstretch{1.125} 
\setlength{\skip\footins}{0.8cm}
\setlength{\footnotesep}{0.25cm}
\setlength{\jot}{10pt}
\titlespacing*{\section}{0pt}{4pt}{4pt}
\titlespacing*{\subsection}{0pt}{15pt}{1pt}

\fancyfoot{}
\fancyfoot[LO,RE]{\vspace{-7.1pt}\includegraphics[height=9pt]{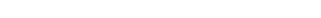}}
\fancyfoot[CO]{\vspace{-7.1pt}\hspace{13.2cm}\includegraphics{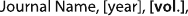}}
\fancyfoot[CE]{\vspace{-7.2pt}\hspace{-14.2cm}\includegraphics{head_foot/RF}}
\fancyfoot[RO]{\footnotesize{\sffamily{1--\pageref{LastPage} ~\textbar  \hspace{2pt}\thepage}}}
\fancyfoot[LE]{\footnotesize{\sffamily{\thepage~\textbar\hspace{3.45cm} 1--\pageref{LastPage}}}}
\fancyhead{}
\renewcommand{\headrulewidth}{0pt} 
\renewcommand{\footrulewidth}{0pt}
\setlength{\arrayrulewidth}{1pt}
\setlength{\columnsep}{6.5mm}
\setlength\bibsep{1pt}

\makeatletter 
\newlength{\figrulesep} 
\setlength{\figrulesep}{0.5\textfloatsep} 

\newcommand{\topfigrule}{\vspace*{-1pt}%
\noindent{\color{cream}\rule[-\figrulesep]{\columnwidth}{1.5pt}} }

\newcommand{\botfigrule}{\vspace*{-2pt}%
\noindent{\color{cream}\rule[\figrulesep]{\columnwidth}{1.5pt}} }

\newcommand{\dblfigrule}{\vspace*{-1pt}%
\noindent{\color{cream}\rule[-\figrulesep]{\textwidth}{1.5pt}} }

\newcommand{\rh}[1]{\textcolor{orange}{(RH:  #1)}}
\newcommand{\changed}[1]{\textcolor{red}{#1}}

\makeatother

\twocolumn[
  \begin{@twocolumnfalse}
{\includegraphics[height=30pt]{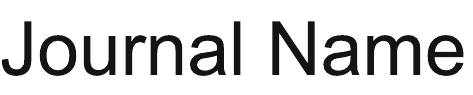}\hfill\raisebox{0pt}[0pt][0pt]{\includegraphics[height=55pt]{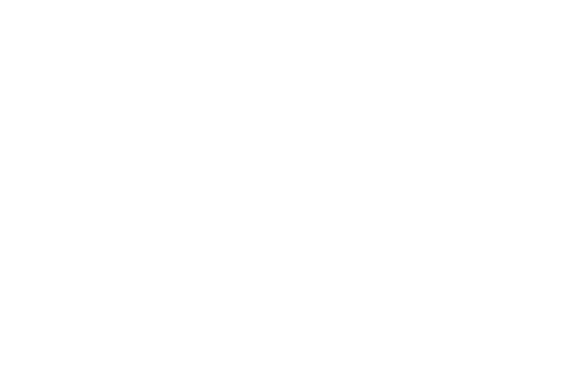}}\\[1ex]
\includegraphics[width=18.5cm]{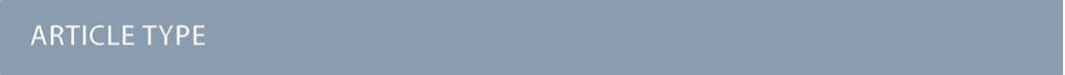}}\par
\vspace{1em}
\sffamily
\begin{tabular}{m{4.5cm} p{13.5cm} }

\includegraphics{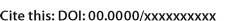} & \noindent\LARGE{\textbf{Denser-than-crystalline disordered polydisperse ellipse packings}} \\
\vspace{0.3cm} & \vspace{0.3cm} \\

 & \noindent\large{Kinsley G. Trlak,\textit{$^{a}$} Dewni S. de Silva,\textit{$^{a}$} Salvatore Torquato\textit{$^{b}$} and Robert S. Hoy\textit{$^{a}$}$^{\dag}$} \\

\includegraphics{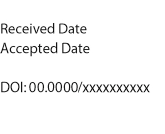} & \noindent\normalsize{
By employing SWAP Monte Carlo moves during slow Lubachevsky-Stillinger-like particle growth, we generate well-thermalized disordered jammed ellipse packings for a wide range of particle aspect ratios $\alpha$ and maximum/minimum particle-diameter ratios $\lambda$.
For $\lambda >  \lambda_{\rm xtal}(\alpha)$, the packings are denser than the perfect monodisperse ellipse crystal ($\phi_{\rm xtal}^{\rm mono} = \pi/2\sqrt{3} \simeq 0.9069$) despite lacking any obvious long-range structural order; we find that $\lambda_{\rm xtal}(\alpha)$ remains below $2.88$ over the range $1 \leq \alpha \leq 5$, and drops below $1.65$ for $\alpha \simeq 1.2$.
These disordered $\phi > \phi_{\rm xtal}^{\rm mono}$ packings are the first of their kind reported for anisotropic particles.
Multiscale analyses reveal that their characteristic structural motifs are strongly $\alpha$- and $\lambda$-dependent, but none of the examined metrics change sharply as the $\lambda = \lambda_{\rm xtal}(\alpha)$ boundary is crossed.
For example, we find that packings with $\alpha \simeq 1.2$ and $\lambda \simeq \lambda_{\rm xtal}(\alpha)$ exhibit a fractionated-hexatic-plastic-polycrystalline structure that (like that of the recently-reported ideal disk glasses) can be exactly-isostatic, triangulated, and effectively-hyperuniform with respect to packing-fraction fluctuations.
In contrast, even-denser packings with larger $\alpha$ and $\lambda$ have none of these properties.
} \\
\end{tabular}

 \end{@twocolumnfalse} \vspace{0.6cm}

  ]

\renewcommand*\rmdefault{bch}\normalfont\upshape
\rmfamily
\section*{}
\vspace{-1cm}


\footnotetext{\textit{$^{a}$~Department of Physics, University of South Florida, Tampa, Florida 33620,USA}}
\footnotetext{\textit{$^{b}$~Department of Chemistry, Department of Physics, and Princeton Institute of Materials, Princeton University, Princeton, New Jersey 08544, USA}}
\footnotetext{$^{\dag}$E-mail: rshoy@usf.edu}



\section{Introduction}
\label{sec:intro}

Any convex shape can form a disordered ``random Apollonian packing'' (RAP) that tiles space (i.e., has a packing fraction $\phi = 1$) in the infinite-polydispersity limit.
Physically-realizable materials, however, are necessarily limited to finite maximum/minimum particle-size ratios $\lambda \equiv \sigma_{\rm max}/\sigma_{\rm min}$.
On the other hand, many real granular materials are \textit{highly} polydisperse.  
For example, materials with power-law-like particle-size distributions $P(\sigma) \sim \sigma^{-\beta}$ over two or more orders of magnitude in $\sigma$ (i.e., materials with $\lambda > 10^2$) include natural systems including  interstellar grains, soils, sea ice, and the aggregates used in the production of concrete.\cite{mathis79,wu93,rothrock84,fuller07}
They also describe, to high precision, the $P(\sigma)$ of various man-made materials generated by explosive and/or maximally-random fragmentation.\cite{turcotte86,villermaux25}

How densely can such materials pack?
Disordered strictly-jammed packings with $\phi > \phi_{\rm xtal}$, where $\phi_{\rm xtal}$ is the crystalline packing fraction, are of longstanding interest because their existence implies that the crystal is not the thermodynamic ground state for hard particles in the infinite-pressure limit.\cite{torquato10}
Such packings have attracted intense interest very recently because (at least for disks) they appear to be the first computationally-produced examples of Kauzmann's ideal glasses.\cite{kauzmann48}
For example, they can have anomalously high bulk moduli, shear moduli, and melting temperatures, and their vibrational density of states $D(\omega)$ exhibits Debye-like scaling in the $\omega \to 0$ limit.\cite{bolton26,fan26}

Simulations can systematically study how particle-size dispersity affects systems' ability to form such packings by employing variable-width $P(\sigma)$.
The $\lambda$- and $\beta$-dependencies of \textit{disk-based} ``physical fractals'' (RAP-like packings with finite $\lambda$) have been carefully investigated.\cite{varrato11,monti23}
Spheres can form $\phi > \phi_{\rm xtal}$ packings whenever $\lambda > \lambda_{\rm min}(\beta,\mathcal{P})$, where $\mathcal{P}$ is a generic set of variables capturing the sample-preparation-protocol dependence.\cite{oquendo20,oquendo22,monti22,monti23,anzivino23,hoy25}
For example, employing SWAP Monte Carlo (SMC) moves\cite{ninarello17,berthier24} during a fast Lubachevsky-Stillinger (LS)-like particle-growth process\cite{lubachevsky91} reduces $\lambda_{\rm min}$ from $\simeq 16$ to $\simeq 5$ for spheres with $\beta = 3$.\cite{hoy25}

Disordered jammed $\phi > \phi_{\rm xtal}$ packings of \textit{anisotropic} particles, however, have not been previously reported.\footnote[1]{Ref.\ \cite{delaney08} examined RAP-like physical ellipse fractals with $\phi > \phi_{\rm xtal}$, but did not generate jammed packings.}
Two-dimensional ellipses are perhaps the best shapes with which to begin a search for such packings because they are a straightforward extension of disks (adding only a single parameter, the aspect ratio $\alpha$) and their jamming phenomemology for preparation protocols which mimic 
\textit{rapid} and/or athermal compression has already been extensively studied.\cite{donev07,delaney05,rocks23,vanderwerf18,jiang26,mailman09,schreck10,schreck12}
Ellipse packings generated in this way share two noteworthy features that are now well understood:
\begin{enumerate}
\item Their jamming densities $\phi_{\rm J}(\alpha)$ first increase with $\alpha$, then reach a maximum at $\alpha = \alpha_{\rm max}$ (with $1.3 \lesssim \alpha_{\rm max}  \lesssim 1.5$ and $0.890  \lesssim \phi_{\rm J}(\alpha_{\rm max})  \lesssim 0.897$), and thereafter decrease steadily.
\item Their coordination numbers $Z_{\rm J}(\alpha)$ first increase in a singular fashion ($Z_{\rm J} - 4 \propto \sqrt{\alpha-1}$), then exhibit a substantially-hypostatic plateau (with $5.5 \lesssim Z_{\rm J} \lesssim 5.8$) at intermediate $\alpha$, and finally decrease steadily for $\alpha \gtrsim 3$.
\end{enumerate}
The initial increase in $\phi_{\rm J}$ arises because ellipses’ ability to rotate away from one another allows them to pack more densely.\cite{donev07,donev04}
Packings remain hypostatic for all $\alpha$ because ellipses can be mechanically stabilized by their curvature at the point of contact \cite{donev07} and/or by quartic vibrational modes.\cite{mailman09,schreck10,schreck12}
Finally, the decreasing $\phi_{\rm J}$ for larger $\alpha$ can be explained using Onsager-like arguments,\cite{onsager49,philipse96} while the decreasing $Z_{\rm J}$ have been associated with the onset of nematic order.\cite{rocks23}

We recently showed\cite{hoy24b} that these trends are \textit{not} universal.
By combining fast LS-like particle growth with (i) \textit{biased} SWAP Monte Carlo (BSMC) moves which swap the diameters of larger particles with smaller interparticle gaps with those of smaller particles with larger interparticle gaps, and (ii) transient degree of freedom (TDOF) moves which allow particles to grow by different amounts,\cite{hagh22,bolton26} we generated disordered ellipse packings with substantially higher ($\phi_{\rm J},Z_{\rm J}$) than were obtained in any previous studies, for a wide range of  $\alpha$.
The $\phi_{\rm J}(\alpha)$ values were within less than 0.5\% of the monodisperse-ellipse crystal packing fraction ($\phi_{\rm xtal}^{\rm mono} = \pi/(2\sqrt{3}) \simeq .9069$\cite{toth50}) over a substantial range ($1.2 \lesssim \alpha \lesssim 1.45$).
Moreover, the intermediate-$\alpha$ packings were (in contrast to those found in all previous studies) \textit{isostatic}.
While the densest packings we reported in Ref.\ \cite{hoy24b} had $\phi < \phi_{\rm xtal}^{\rm mono}$ (specifically, $\phi_{\rm J}(\alpha_{\rm max}) = 0.9044$), the small value of $\lambda$ employed in this study ($\lambda = 1.4$) suggested that disordered $\phi > \phi_{\rm xtal}^{\rm mono}$ packings of modestly-polydisperse ellipses can be produced using a comparable protocol.
In this paper, we show that indeed they can.

\begin{figure}[h]
\centering
\includegraphics[width=3.25in]{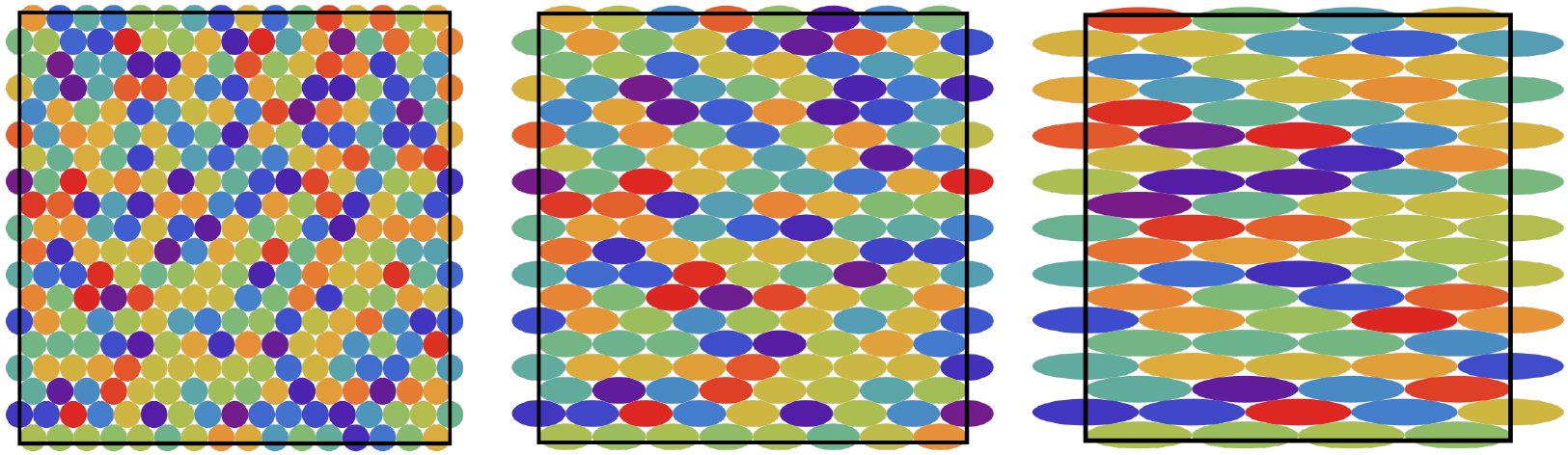}
\includegraphics[width=3.25in]{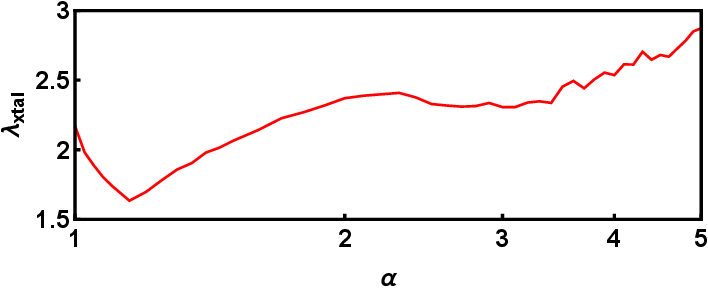}
\caption{The top panels show monodisperse ellipse crystals for $\alpha = 1$, $2$ and $4$, while the bottom panel shows our results for $\lambda_{\rm xtal}(\alpha)$. 
Systems with $\lambda > \lambda_{\rm xtal}(\alpha)$ can form disordered packings with $\phi_{\rm J}(\alpha) > \phi_{\rm xtal}^{\rm mono}$.} 
\label{fig:xtal}
\end{figure}

Before proceeding further, we clarify what we mean by ``denser than crystalline''.
Toth proved\cite{toth50} that the densest monodisperse ellipse crystals are affinely-stretched hexagonal lattices such as those illustrated in Figure \ref{fig:xtal}.
These crystals have $\phi = \phi_{\rm xtal}^{\rm mono}$,  $Z = 6$, and perfect nematic order.
Continuously-polydisperse ellipses can always form infinitely-fractionated locally-monodisperse crystals with $\phi = \phi_{\rm xtal}^{\rm mono}$ in the thermodynamic limit, but such crystals cannot be the infinite-pressure thermodynamic ground states for any ($\alpha,\beta,\lambda$) for which disordered packings with $\phi > \phi_{\rm xtal}^{\rm mono}$ exist.
The densest crystalline packings of bidisperse ellipses remain unknown in general, 
but one representative class is formed by affinely stretching a bidiperse crystal wherein the small disks exactly fill the gaps in a maximally-dense hexagonal lattice formed by the large disks.
Such crystals have $\lambda = \lambda_{\rm crit} = 3 + 2\sqrt{3} \simeq 6.46$, $\phi \equiv \phi_{\rm xtal}^{\rm bi} = 0.9503$ and $Z = 6$.\cite{uche04}

Here we focus on $\lambda$ that are well below $\lambda_{\rm crit}$, where formation of infinitely-fractionated locally-bidisperse crystals with the abovementioned structure is impossible.
By combining \textit{slow} LS-like particle growth with SWAP and TDOF moves, we generate disordered $\phi > \phi_{\rm xtal}^{\rm mono}$ packings for all $1 \leq \alpha \leq 5$ and all $\lambda \geq \lambda_{\rm xtal}(\alpha)$ [Fig.\ \ref{fig:xtal}].
By performing a wide range of analyses of their multiscale structure, we find that the structure of these packings does \textit{not} change noticeably as the $\lambda = \lambda_{\rm xtal}(\alpha)$ boundary is crossed.
On the other hand, their intermediate-range order varies strongly and nontrivially with both $\alpha$ and $\lambda$, and strikingly-different local motifs are dominant in different ($\alpha,\lambda$)-regimes.

For $\alpha \simeq \alpha_{\rm max} \simeq 1.2$, we find $\phi > \phi_{\rm xtal}^{\rm mono}$ packings for $\lambda$ values as low as $\lambda_{\rm xtal}^{\rm min} \simeq 1.63$.
For $\lambda \simeq \lambda_{\rm xtal}^{\rm min}$, these packings' stucture necessarily differs greatly from both the $\lambda \to 1$ monodisperse-crystal and $\lambda \to \infty$ RAP-like limits.
We find that such packings are distinguished by their combination of moderate fractionation and a high degree of local hexatic order with minimal intermediate-range nematic order and packing-fraction fluctuations.
Some of them have three additional properties that are shared by their  ideal-glass-like-disk-packing counterparts;\cite{bolton26,fan26} specifically, they are exactly-isostatic, triangulated and effectively-hyperuniform.\cite{torquato03,torquato18b,chen18}
For the same aspect ratios but larger $\lambda$, we find even-denser packings with two characteristic structural motifs: (i) the largest particles are often surrounded by at least twelve small neighbors, and (ii) the smallest particles are often trapped between three large neighbors.
The same motifs also appear in systems where both $\alpha$ and $\lambda$ are large.
These motifs are also present in the abovementioned $\lambda = \lambda_{\rm crit}$ bidisperse crystals,\cite{uche04} but here they appear within packings that that lack any obvious long-range structural order.

For larger aspect ratios, the structure of jammed packings with $\lambda > \lambda_{\rm xtal}(\alpha)$ arises in part from the fact that two of Ref.\ \cite{hoy24b}'s main results are not specific to small $\lambda$.
First, well-thermalized particle growth (in contrast to poorly-thermalized growth\cite{donev07,delaney05,rocks23,vanderwerf18,jiang26}) produces remarkably-flat $\phi_{\rm J}(\alpha > \alpha_{\rm max},\lambda)$ for all $1.4 \lesssim \lambda \leq 5$ because it allows generation of jammed states with a high degree of intermediate-range nematic order.
The onset of this ordering produces the maximum in $\lambda_{\rm xtal}(\alpha)$ at $\alpha \simeq 2.3$.
We find that increasing $\lambda$ does not suppress this order, and indeed that $\phi_{\rm J}(\alpha,\lambda)$ increases monotonically with $\lambda$ for all $\alpha \gtrsim \alpha_{\rm max}$.
Second, well-thermalized particle growth (in contrast to poorly-thermalized growth \cite{donev07,delaney05,rocks23,vanderwerf18}) produces nearly-isostatic packings over a broad range of $\alpha$.
We find that $Z_{\rm J}(\alpha,\lambda)$ is only weakly $\lambda$-dependent for $\lambda \gtrsim 1.4$.
On the other hand, its standard deviation $\Delta Z_{\rm J}(\lambda)$ increases rapidly with $\lambda$.
We find that $\Delta Z_{\rm J}(\alpha,\lambda) - \Delta Z_{\rm J}(\alpha_{\rm max},\lambda) -$ correlates strongly with $\phi_{\rm J}(\alpha_{\rm max},\lambda) - \phi_{\rm J}(\alpha,\lambda)$ and the magnitude of intermediate-range packing-fraction fluctuations, but only for $\alpha \simeq \alpha_{\rm max}$ and sufficiently-small $\lambda$.
For larger $|\alpha - \alpha_{\rm max}|$ and $\lambda$, broad $P(Z_i)$, where $Z_i$ is particle $i$'s coordination number, appear to offer systems more flexibility to form stable ultradense packings.

The outline of the rest of this paper is as follows:
Section \ref{sec:methods} describes our packing-generation protocol and the metrics we use to analyze these packings' multiscale structural order.
Section \ref{sec:results} presents the results of these analyses and compares them to those obtained in previous studies of $\phi < \phi_{\rm xtal}$ packings.
Section \ref{sec:conclude} places these results in context with each other and with previous work, and presents our conclusions.
Finally, an Appendix provides additional details.

\section{Methods}
\label{sec:methods}

\subsection{Generating jammed packings}
\label{subsec:IIA}

As in Refs.\ \cite{rocks23,hoy24b}, we begin by placing $N$ nonoverlapping ellipses of aspect ratio $\alpha$, with random positions and orientations, in square $L \times L$ domains with $L = \sqrt{N\alpha}$.
Periodic boundary conditions are applied along both directions.
We employ a power-law distribution for the initial minor-axis lengths
\begin{equation}
P_\lambda(\sigma) = \Bigg{ \{ } \begin{array}{ccl}
\displaystyle\frac{\lambda }{\lambda-1}  \displaystyle\frac{\sigma_{\rm min}}{\sigma^2} & \ \ , \ \ & \sigma_{\rm min} \leq \sigma \leq \lambda\sigma_{\rm min} \\
\\
0 & \ \ , \ \ & \sigma < \sigma_{\rm min}  \ \rm{or}\ \sigma > \lambda\sigma_{\rm min} 
\end{array}.
\label{eq:Pofsigma}
\end{equation}
The initial value of $\sigma_{\rm min}$ is chosen to obtain an initial packing fraction $\phi_{\rm init} \simeq 0.01$ that is low enough to eliminate most $\phi_{\rm init}$-dependent effects.\cite{chaudhuri10,ozawa17,monti25}

This $P(\sigma)$ has been employed in multiple recent studies of disk\cite{monti23,kim24,fan26} and ellipse\cite{rocks23,hoy24b} jamming.
It appears to optimize glass-formation at moderate $\lambda$,\cite{ninarello17} and also corresponds to a flat distribution of particle areas.
The fractal dimension associated with the $\lambda \to \infty$ limit of this distribution ($d_{\rm{f}} \equiv \beta - 1 = 1$) is comparable to, but significantly lower than, the values for random Apollonian disk packings ($d_{\rm{f}}^{\rm{Apoll}} = 1.305$\cite{varrato11}) and packings of particles produced by maximally-random fragmentation $d_{\rm{f}}^{\rm{frag}} \simeq 1.4$\cite{villermaux25}).
In other words, we did \textit{not} utilize a $\beta$ value that was chosen to optimize $\phi_{\rm J}$.

Next we begin the particle-growth procedure. 
Thoughout the entire process described above and below, overlaps between ellipse pairs $(i,j)$ are prohibited using Zheng's exact expression\cite{zheng07} for their orientation-dependent distance of closest approach $d_{\rm cap}(i,j)$.
The procedure consists of three stages per growth cycle:
\begin{enumerate}
\item Attempting to translate each particle $i$ by a random displacement along each Cartesian direction and rotate it  by a random angle.
\item Increasing all particles' minor-axis lengths $\sigma$ by the \textit{same} factor $\mathcal{F} = \min(10^{-\Delta}/(2\phi),\tilde{G})$, where $\tilde{G}$ is the value that brings \textit{one} pair of ellipses into tangential contact.\cite{rocks23,hoy24b}
\item $N$ attempts to SWAP the minor-axis lengths of particles $i$ and $j$ that have $4/5 \leq \sigma_i/\sigma_j \leq 5/4$\cite{ninarello17} but are otherwise randomly selected.
\end{enumerate}
Stage (1) is implemented in a manner similar to that described in Refs.\ \cite{rocks23,hoy24b}, but with three refinements aimed at improving thermalization.
First, the maximum translational-displacement magnitudes are scaled by $\langle\sigma\rangle/\sigma_i$ since particles are assumed to have kinetic energies $m_i v_i^2/2 \sim \sigma_i^2 v_i^2 \sim k_BT$.\footnote[2]{Here the units of the speed $v$ and temperature $T$ are arbitrary since the unit of time is undefined.}
Second, the maximum rotational-displacement magnitudes are scaled by $(\langle\sigma\rangle/\sigma_i)^2/\sqrt{1 + \alpha^2}$ since particles' moments of inertia are $I_i = m_i\sigma_i^2(1 + \alpha^2)$.
Third, instead of making at most one translation/rotation attempt per particle, attempts are continued until either one succeeds or 100 have failed.
Since this repetition of move attempts can cause violations of detailed balance, our protocol should (strictly speaking) be regarded as a packing-optimizer rather than as a method of generating equilibrium states.

Stage (2) increases $\phi$ by a maximum of $10^{-\Delta}$; smaller increments (i.e., $2\phi \tilde{G}$) occur when they are needed to avoid introducing interparticle overlaps.
Previous comprehensive studies of ellipse jamming, including very recent studies like Refs.\ \cite{rocks23,hoy24b,jiang26}, have all employed fast particle growth.
Here we aim to improve thermalization and maximize $\phi_{\rm J}$ by employing the slowest particle growth that allows $N = 10^4$ packing-generation runs to finish within a reasonable amount of time (i.e., one month) even for large $\alpha$ and $\lambda$.
Therefore we set $\Delta$ as follows: $\Delta = 4$ for $\phi < 0.4$, $\Delta = 5$ for $0.4 \leq \phi < 0.6$,  $\Delta = 6$ for $0.6 \leq \phi < 0.8$, and  $\Delta = 7$ for $\phi \geq 0.8$.
This increasingly-restrictive capping of the maximum particle-growth rate allows systems to spend much more time at higher $\phi$, where their relaxation dynamics are far slower.\cite{letz00,pfleiderer08,zheng11,mishra13,xu15}

In the early stages of this study, we found that the BSMC algorithm employed in Ref.\ \cite{hoy24b} only increases $\phi_{\rm J}$ when fast particle growth is employed.
Therefore, since the standard implentation of the SMC algorithm is known to generate equilibrium configurations over a very wide range of $\phi$,\cite{berthier24} we employed standard SMC in stage (3).
An additional advantage of standard SMC over BSMC is that its associated computational effort is $\mathcal{O}(N)$ rather than $\mathcal{O}[N\ln(N)]$ or worse, allowing larger packings to be produced.
Note that the restriction on $\sigma_i/\sigma_j$ was necessary to obtain reasonable SWAP success rates for larger $\lambda$.\cite{ninarello17}

We manage these packing-generation runs using a refined version of the protocol employed in Ref.\ \cite{hoy24b}.
The attempted translations and rotations during stage (1) have maximum magnitudes $|\delta\vec{r}_i| = 0.05f \langle\sigma\rangle/\sigma_i$ and $|\delta\theta_i| = [16f(\langle\sigma\rangle/\sigma_i)^2/\sqrt{1 + \alpha^2}]^\circ$, respectively.
Here the move-size factor $f$ is set to $1$ at the beginning of all runs, and multiplied by $1/2$ whenever $100$ consecutive growth cycles have passed with $\tilde{G} < 10^{-8}$.
When $f$ drops below $10^{-2}$, we change stage (2) to
\begin{itemize}
\item [2b.] Increasing each particle's minor-axis length by a factor $G_i/2$, where $G_i$ is the factor that would bring it into contact with its closest neighbor.\cite{zheng07}
\end{itemize}
Since stage (2b) allows the shape of $P(\sigma)$ to vary, and effectively adds one degree of freedom per particle, we refer to these increases as TDOF moves.
We find, however, that since here they are only implemented very late in the runs (i.e., after systems are already almost jammed), they only produce minimal changes in the shape of $P(\sigma)$, especially in comparison to the moves employed in Refs.\ \cite{hagh22,bolton26,fan26}.
The primary effect of the TDOF moves employed in \textit{this} study is allowing rattlers to continue growing even after the rest of the system has formed a jammed backbone.\cite{moukarzel98}
While this process leads to only minimal increases in $\phi_{\rm J}$, it can produce substantial increases in $Z_{\rm J}$ that become more dramatic with increasing $\lambda$.\cite{hoy25}

The main growth procedure terminates when $\max(|\delta\vec{r}_i|) = 0.05f \langle\sigma\rangle/\sigma_{\rm min}$ drops below $10^{-10}$, which is the smallest value allowed by our double-precision numerical implementation.
Stage (2b) is then repeated until all particles have $G_i = 0$.
When this occurs, the system is considered jammed and the packing-generation run terminates.

\subsection{Analyzing their structural order}
\label{subsec:IIB}

Following multiple previous studies,\cite{donev07,schreck10,schreck12,rocks23,hoy24b} we examine the jammed packings' coordination-number statistics.
The average coordination number is simply defined as $Z_{\rm J} = \langle Z_i \rangle$, where $Z_i$ is the number of interparticle contacts including particle $i$.
Since overlaps are non-existent in hard-particle packings, these contacts must be defined using a finite numerical tolerance $C$.
Here we define $Z_i$ using the commonly-employed relation $Z_i = \sum_{j \neq i} \Theta\left([d_{\rm cap}(i,j) + C] - r_{ij}\right)$, where $\Theta$ is the Heaviside step function.
All results presented below are for $C = 10^{-3}$, which lies well within $Z_{\rm J}(C)$'s plateau regime.\cite{donev05c}
We also present results for the standard deviation of this quantity, i.e.\ $\Delta Z_{\rm J} = \sqrt{\langle Z_i^2 \rangle - \langle Z_i \rangle^2}$.
This quantity does not seem to have been investigated in previous work.
Also note that in contrast to some previous studies,\cite{donev07,schreck10,schreck12} we do \textit{not} make any attempt to identify and remove rattlers before calculating the $\{ Z_i \}$.
Nor do we attempt to account for finite-$N$ corrections to isostaticity, simulation-cell degrees of freedom, or ``dependent'' constraints.\cite{donev07}

Local hexatic order is investigated using the metric
\begin{equation}
\Psi_6(i) = \displaystyle\frac{1}{6} \displaystyle\sum_{j = 1}^6 \exp(6\mathrm{i}\zeta_{ij}).
\label{eq:Psi6ofi}
\end{equation}
Here $\zeta_{ij}$ is the angle between the vector $\vec{r}_{ij} \equiv \vec{r}_j - \vec{r}_i$ and a reference direction, while the sum is taken over the six nearest neighbors of particle $i$.
Note that $\langle | \Psi_6(i) | \rangle \geq |\langle  \Psi_6(i)  \rangle |$, where the latter quantity is the standard global-hexatic-order metric
\begin{equation}
\Psi_6 = \bigg{|}  \displaystyle\frac{1}{N} \displaystyle\sum_{i=1}^{N} \Psi_6(i) \bigg{|}.
\label{eq:standardHOP}
\end{equation}
In the $N \to \infty$ limit, $\langle \Psi_6 \rangle$ is $1$ ($0$) for a system with perfect hexatic order (no orientational order).

Local nematic order is investigated using the orientational-order metric
\begin{equation}
S_{\rm loc}(\mathcal{N}) = \bigg{\langle}  \displaystyle\frac{1}{\mathcal{N}} \displaystyle\sum_{j = 1}^\mathcal{N}\left(2\cos^2(\Delta \theta_{ij}) - 1 \right) \bigg{\rangle},
\label{eq:SofN}
\end{equation}
where $\Delta \theta_{ij} \equiv \theta_j - \theta_i$ is the orientation-angle difference between ellipses $i$ and $j$, the sum is taken over the $\mathcal{N}$ nearest neighbors of particle $i$, and the average is taken over all $i$.
This quantity is a local version of the standard 2D global-nematic-order parameter
\begin{equation}
S_{\rm glob}(N)= \langle 2\cos^2(\theta) -1 \rangle   \equiv \displaystyle\frac{1}{N} \displaystyle\sum_{i=1}^N \left(2\cos^2(\theta_i) - 1 \right)
\label{eq:P2ofN}
\end{equation}
that has been adapted to make $\theta_i$ define the local reference direction, and then spatially averaged. 
In the thermodynamic limit, $S_{\rm loc}(\mathcal{N})$ is $1$ ($0$) for a system with perfect nematic order (no orientational order).

Packing-fraction fluctuations on the same length scale are examined using the local packing fractions $\{ \phi_i(\mathcal{N}) \}$, where $\phi_i(\mathcal{N}$) is the packing fraction within the circular domain of radius $R = \sqrt{\alpha \mathcal{N}/\pi}$ that is centered on $\vec{r}_i$.
These domains include approximately $\mathcal{N}$ ellipse centers, and we calculate the  $\phi_i$ by averaging over $n_{\rm samp} = 10^4\mathcal{N}$ randomly selected points within each domain.
The associated packing-fraction-fluctuation metric is
\begin{equation}
\delta \phi(\mathcal{N}) =\sqrt{\langle \phi_i^2(\mathcal{N}) \rangle - \langle \phi_i(\mathcal{N}) \rangle^2},
\label{eq:phiflucts1}
\end{equation}
where angle brackets indicate averaging over all $i$.
Ref.\ \cite{hoy24b} showed that $\delta \phi(\mathcal{N})$ is minimized at $\alpha \simeq \alpha_{\rm max}$ for fast LS-like growth and $\lambda = 1.4$, for various $\mathcal{N} \leq 100$.
Here we will focus on results for $\mathcal{N} = 100$.

Finally, to characterize the packings' hyperuniformity or lack thereof, we investigate their spectral densities $\hat{\chi}_{\rm V}(k)$ using the method described in Ref.\ \cite{zachary11c}.
By defining ``geometric parameters'' $\tilde{r}_i$ for each particle $i$ (using its center position $\vec{r}_i$, orientation $\theta_i$ and size $\sigma_i$), one obtains
\begin{equation}
\begin{array}{c}
\hat{\chi}_{\rm V}(\vec{k}) = L^{-2} \bigg{|}  \displaystyle\sum_{i = 1}^N  \exp(-\textrm{i}\vec{k} \cdot \vec{r}_i)  \hat{m}_{\rm V}(\vec{k}; \tilde{r}_i) \bigg{|}^2 , \\
\\
\textrm{for}  \ \ \vec{k} = \displaystyle\frac{2\pi}{L} \left( n_x \hat{x} + n_y \hat{y} \right),  \ \ \rm{with}  \ \  (n_x, n_y) \in \mathbb{Z}^2.
\end{array}
\label{eq:ChiOfK1}
\end{equation}
Here $\hat{m}_{\rm V}(k)$ is the Fourier transform of the particle volume indicator function $m_{\rm V}(\vec{r}; \tilde{r}_i)$.
This function is defined so that $m_{\rm V}(\vec{r}; \tilde{r}_i) = 1$ ($0$) when $\vec{r}$ is inside (outside) particle $i$.\cite{torquato18b,zachary11c}

\section{Results}
\label{sec:results}

Unless otherwise specified, all results presented below are for single $N = 10^4$ packings.
The associated periodic simulation-cell side lengths $L$, which are approximately $\sqrt{N/(\alpha\lambda)}$ times larger than the largest ellipses' major-axis lengths, are sufficient to minimize finite-size effects.

\begin{figure}
\centering
\includegraphics[width=3in]{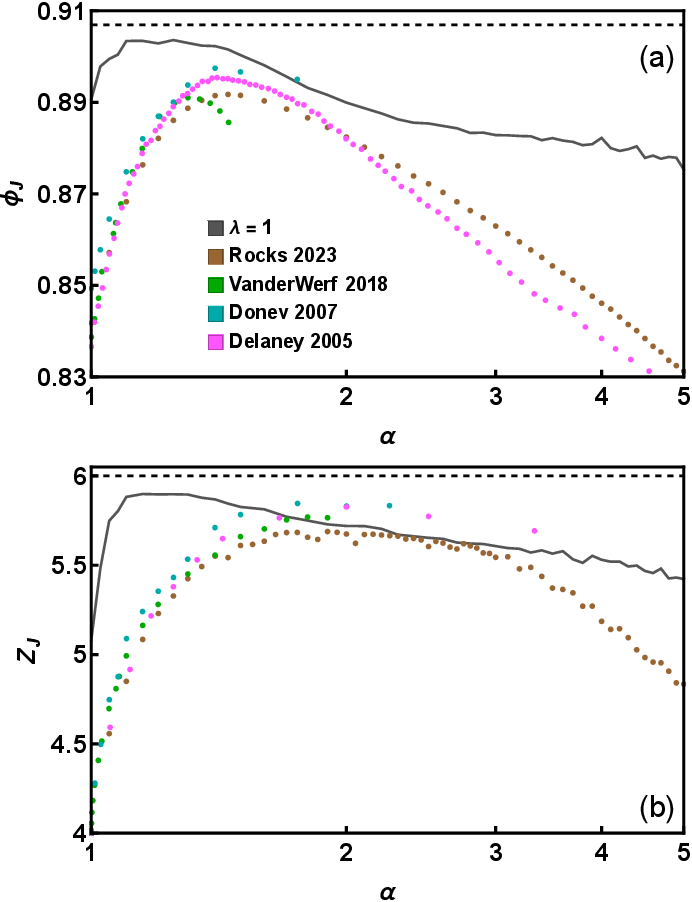}
\caption{Comparison of the $\phi_{\rm J}(\alpha)$ and $Z_{\rm J}(\alpha)$ we obtained for $\lambda = 1$ to results from previous studies of ellipse jamming that employed $\lambda \leq 1.4$ and fast particle growth.\cite{delaney05,donev07,vanderwerf18,rocks23}   Dashed lines respectively indicate $\phi = \phi_{\rm xtal}$ and $Z_{\rm J} = Z_{\rm iso}$.}
\label{fig:phiandZ2}
\end{figure}

To set the stage for the remainder of this section, we will begin our discussion by comparing some of our results for $\lambda = 1$ to those obtained in previous studies of ellipse jamming that employed $\lambda \leq 1.4$ and fast particle growth.\cite{delaney05,donev07,vanderwerf18,rocks23}
Figure \ref{fig:phiandZ2}(a) shows that our results for $\phi_{\rm J}(\alpha)$ follow the same qualitative trends observed in previous work.
Specifically, $\phi_{\rm J}$ increases linearly with $\alpha$ for $\alpha - 1 \ll 1$, reaches a peak at $\alpha = \alpha_{\rm max}$, and thereafter decreases steadily.
On the other hand, there are three quantitative differences resulting from the slower particle growth (Sec.\ \ref{subsec:IIA}) employed here.
First and second, $\alpha_{\rm max} \simeq 1.25$ is slightly lower, and $\phi_{\rm J}(\alpha_{\rm max})= 0.9036$ is substantially higher, than the values obtained using fast particle growth ($1.3 \lesssim \alpha_{\rm max} \lesssim 1.45$ and  $0.891 \lesssim \phi_{\rm J}(\alpha_{\rm max}) \lesssim 0.897$).
These differences arise primarily from the higher level of hexatic order enabled by the combination of slow particle growth and small $\alpha$.
Third, $\phi_{\rm J}$ decreases substantially slower with increasing $\alpha$, particularly for $\alpha > 2$.
This arises primarily from the higher level of nematic order enabled by the combination of slow particle growth and large $\alpha$.

Fig.\ \ref{fig:phiandZ2}(b) shows that our results for $Z_{\rm J}(\alpha)$ differ from those obtained in previous work in two significant ways.
First, previous studies found $\lim_{\alpha \to 1} Z_{\rm J}(\alpha) = Z_{\rm J}(1) + \mathcal{O}(\sqrt{\alpha-1})$, with $Z_{\rm J}(1) = 4$ corresponding to isostatic disk packings.
Here we find the same $\mathcal{O}(\sqrt{\alpha-1})$ scaling (to within our statistical uncertainties), but $Z_{\rm J}(1) \simeq 5.08$.
This much-higher value of  $Z_{\rm J}(1)$ arises from the combination of the greater crystalline order enabled by the slower particle-growth rate\cite{lubachevsky91} and extra contacts added by the TDOF moves applied at the end of the packing-generation runs.\cite{hoy24b,hoy25}
Second, previous studies found that  $Z_{\rm J}(\alpha)$ continues increasing as $\alpha$ increases beyond the square-root-scaling regime, with decreasing $\partial Z_{\rm J}/\partial \alpha$, until it levels out onto a plateau over the range $1.5 \lesssim \alpha \lesssim 3$.
Here we find that $Z_{\rm J}$ quickly reaches a maximum ($Z_{\rm J}^{\rm max} \simeq 5.90$) at $\alpha \simeq \alpha_{\rm max}$, and thereafter decreases monotonically, albeit slowly, with increasing $\alpha$.

The differences described above, while significant, are minor in comparison to those we will describe below.
In particular, all of our $\lambda = 1$ systems have $\phi_{\rm J}$ that are well below $\phi_{\rm xtal}$ and $Z_{\rm J}$ that are well below $Z_{\rm iso} \equiv 6$.
Two factors allow for much greater enhancements to both $\phi_{\rm J}$ and $Z_{\rm J}$ when $\lambda > 1$.
First, SWAP moves speed up systems' dynamics by orders of magnitude,\cite{berthier16b,ninarello17} allowing systems to bypass the kinetic traps that would otherwise prevent them from maximizing $\phi_{\rm J}$ and $Z_{\rm J}$.\cite{maher21,hoy24b}
Second, ``sufficiently'' large $\lambda$ allow more-efficient packing by allowing smaller particles to fill in the gaps between the larger particles.
In the following sections, we will explore these effects in detail, focusing on results for $\lambda > 1$.

\begin{figure*}
\centering
\includegraphics[width=7in]{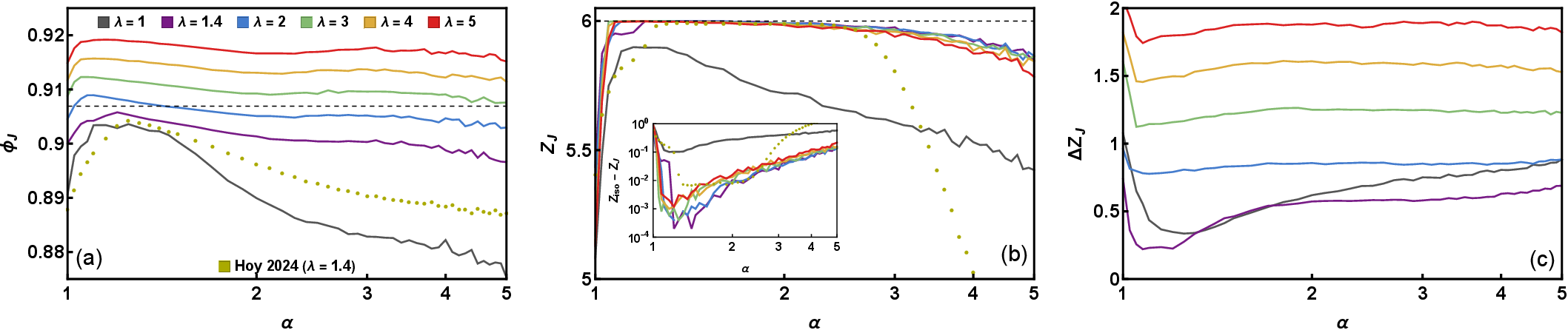}
\caption{Variation of jamming densities [panel (a)], average coordination numbers  [panel (b)], and standard deviations of the coordination numbers [panel (c)]  with $\alpha$ for selected $\lambda$.  
Dashed lines in panels (a-b) respectively indicate $\phi = \phi_{\rm xtal}$ and $Z_{\rm iso} \equiv 6$.
Symbols in these panels show results from our previous study of $\lambda = 1.4$ ellipse packings,\cite{hoy24b}  which  also utilized SWAP and TDOF moves but employed \textit{fast} particle growth.
The inset to panel (b) highlights the degree of hypostaticity $Z_{\rm iso} - Z_{\rm J}$.}
\label{fig:phiandZ1}
\end{figure*}

\subsection{Jamming densities and coordination numbers}
\label{subsec:IIIA}

Figure \ref{fig:phiandZ1} shows the variation of $\phi_{\rm J}$, $Z_{\rm J}$ and $\Delta Z_{\rm J}$ with $\alpha$ for selected $\lambda$.
The variation of these quantities with $\lambda$ for selected $\alpha$ is discussed in the Appendix.

Panel (a) shows that $\phi_{\rm J}$ increases monotonically with $\lambda$ for all $1 \leq \alpha \leq 5$ (at least for the $\{ \lambda \}$ shown here).
A simple explanation for the positive $\partial \phi_{\rm J}/\partial \lambda$ is as follows:
For ``small'' $\lambda$, broadening $P(\sigma)$ allows systems to form a wider variety of maximally-dense \textit{intermediate-scale} structures.
While such structures are usually geometrically-frustrated and hence cannot tile space, broadening $P(\sigma)$ also allows systems to form a wider variety of efficiently-packed structures in the regions lying between them.
For ``large'' $\lambda$, the positive $\partial \phi_{\rm J}/\partial \lambda$ arise from increasing the ability of smaller particles to efficiently fill in the gaps between the larger ones.
This effect should increasingly dominate packings' structure for $\lambda \gtrsim \lambda_{\rm crit}$, and persist all the way to the $\lambda \to \infty$ RAP-like limit.
As we will discuss below, in practice the above combination of cooperating/competing processes produces strongly and nontrivially $\alpha$- and $\lambda$-dependent intermediate-scale order.

Panel (a) also shows that the $\phi_{\rm J}(\alpha)$ curves rapidly ``flatten out'' with increasing $\lambda$.
In sharp contrast to both our results for $\lambda = 1$ and previous results for fast particle growth,\cite{delaney05,donev07,vanderwerf18,rocks23,jiang26}  
$\phi_{\rm J}$ varies over a range of less than $1.1\%$ for all $\lambda \geq 1.4$, and less than $0.5\%$ for all $\lambda > 3$, over the entire range of $\alpha$ studied here.
The magnitude of the increases in $\phi_{\rm J}$ over the range $1 \leq \alpha \leq \alpha_{\rm max}(\lambda)$ and the decreases in $\phi_{\rm J}$ over the range $\alpha_{\rm max}(\lambda) \leq \alpha \leq 5$ both rapidly diminish with increasing $\lambda$.
Indeed, our $\lambda = 5$ systems actually pack slightly more densely for $\alpha = 5$ ($\phi_{\rm J} = 0.9152$) than they do for $\alpha = 1$ ($\phi_{\rm J} = 0.9149$).
This is surprising, as it indicates that the SWAP-induced dynamical speedup (when combined with slow particle growth) is so effective in promoting increased intermediate-range nematic order that it essentially eliminates Onsager-type effects.\cite{onsager49,philipse96}

Panel (b) shows that the combination of SWAP and TDOF moves with slow particle growth yields packings that are nearly isostatic for a very wide range of $\alpha$ and $\lambda$.
$Z_{\rm J}$ increases more rapidly with $\alpha$ for $\lambda > 1$ than it does for $\lambda = 1$ owing to SWAP's ability to generate locally-optimized structure.\cite{ozawa17}
For all $\lambda \geq 2$, $Z_{\rm J}$ is within $0.03\%$ of $Z_{\rm iso}$, i.e.\ $Z_{\rm J} \geq 5.998$, by $\alpha = 1.1$.
This nearly-isostatic character persists over a broad range of aspect ratios; $Z_{\rm J}$ remains within less than $0.5\%$ of $Z_{\rm iso}$ (i.e., $Z_{\rm J} \geq 5.97$) up to $\alpha > 2.5$ for all $\lambda > 1$.
$Z_{\rm J}$ begins decreasing substantially with increasing $\lambda$ only for $\alpha > 2.5$.
It does so more rapidly for larger $\lambda$, but $Z_{\rm J}$ remains well above $5.8$ all the way to $\alpha = 5$ for all $\lambda < 5$.
Notably, as illustrated by the comparisons to the data from Ref.\ \cite{hoy24b}, the abovementioned decreases in $\phi_{\rm J}$ and $Z_{\rm J}$ are both strongly suppressed by employing slow  particle growth.

The inset to panel (b) shows that $Z_{\rm J}$ comes within less than $10^{-3}$ of $Z_{\rm iso}$ over a wide range of $\alpha$ and $\lambda$.
As mentioned above, we found a few ($\alpha,\lambda$) combinations [not shown here] that yield exactly-isostatic packings with $Z_{\rm J} = Z_{\rm iso}$; these packings will be discussed in Section \ref{subsec:IIID}.
However, we did \textit{not} find any hyperstatic ($Z_{\rm J} > Z_{\rm iso}$) packings.
Ellipses appear to be unable to form such packings, despite their ability to have individual $Z_i$ as high as $16$ [as we observed for $(\alpha,\lambda) = (3.8,5)$].

Panel (c) illustrates how $\Delta Z_{\rm J}$ depends on $\alpha$ and $\lambda$.
For all $\alpha$, the $Z_i$-distributions grow narrower  over the range $1 \leq \lambda \lesssim 1.4$, and then grow broader as $\lambda$ continues increasing.
Surprisingly, $\Delta Z_{\rm J}$ increases rapidly over the range $1.4 \lesssim \lambda \leq 5$ even though $Z_{\rm J}$ is nearly $\lambda$-independent over the same range.
This suggests that the increasingly-diverse locally-optimized structures that enable the positive $\partial \phi_{\rm J}/\partial \lambda$ illustrated in panel (a) are made possible (in part) by the ability to have wider $Z_i$-distributions while retaining $\langle Z_i \rangle \simeq Z_{\rm J} \simeq 6$ above some critical length scale.
On the other hand, this increasing diversity decreases $\phi_{\rm J}$ for fixed $\lambda$; we find that $\Delta Z_J$ is minimized at $\alpha \simeq \alpha_{\rm max}(\lambda)$ for all $\lambda$.

\begin{figure}[h]
\centering
\includegraphics[width=3in]{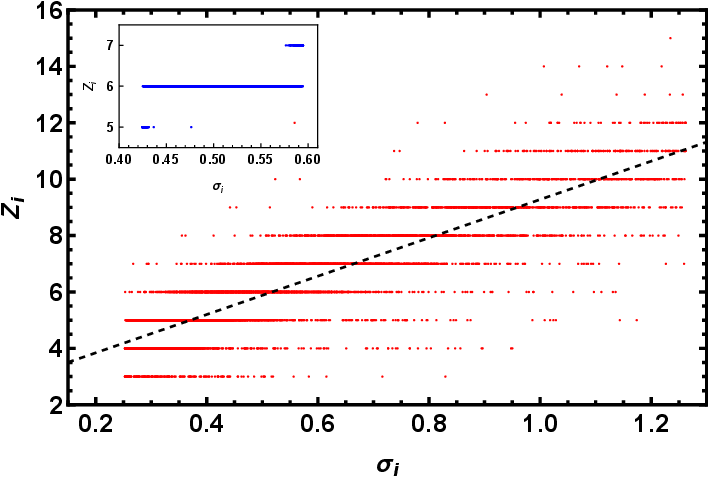}
\caption{Scatter plot of $Z_i$ vs.\ $\sigma_i$ for $(\alpha,\lambda) = (5,5)$.  The dashed line shows $Z = 2.484 + 6.795\sigma$, while the inset shows results for  $(\alpha,\lambda) = (1.2,1.4)$.}
\label{fig:ZiVsSigmaI}
\end{figure}

We also find that while particles with fixed $\sigma_i$ can have a wide range of $Z_i$ values and vice versa, the two quantities are nonetheless strongly correlated.
Figure \ref{fig:ZiVsSigmaI} shows the $\{ (\sigma_i,Z_i) \}$ for  $(\alpha,\lambda) = (5,5)$ along with a linear fit to this data.
As mentioned in the Introduction, the strong correlation of $Z_i$ with $\sigma_i$ leads to two distinctive structural motifs (to be explored  further in Section \ref{subsec:IIIE}) in  packings with sufficiently large $\lambda$.
Note that the degree of scatter in such plots decreases monotonically with decreasing $\lambda$ over the range $5 \geq \lambda \gtrsim 1.4$, while for fixed $\lambda$ it decreases with decreasing $\Delta Z_{\rm J}$ and increasing $\phi_{\rm J}$.
The inset illustrates the extreme case $(\alpha,\lambda) = (1.2,1.4)$.
All particles in this system have $5 \leq Z_i \leq 7$, and almost 95\% of the particles have $Z_i = 6$.

\subsection{Average structural order}
\label{subsec:IIIB}

\begin{figure*}
\centering
\includegraphics[width=7in]{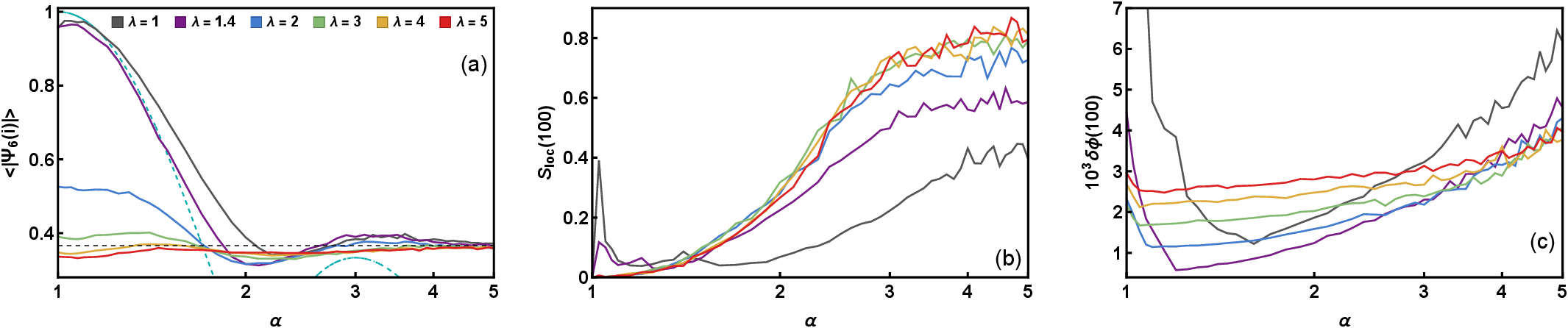}
\caption{Panel (a): Average single-particle hexatic order $\langle|\Psi_6(i)|\rangle$ for the systems highlighted in Fig.\ \ref{fig:phiandZ1}.  The dashed line shows $\langle|\Psi_6|\rangle_{\rm rand}$, while the dashed curve shows  $\langle|\Psi_6(i)|\rangle$ for affinely-stretched hexagonal lattices like those illustrated in Figure \ref{fig:xtal}.
Panel (b): Average intermediate-range nematic order $S_{\rm loc}(100)$ for these systems.
Panel (c): Average intermediate-range packing-fraction fluctuations $\delta\phi(100)$ for these systems.
Note that the lowest reported values are roughly three times the estimated statistical noise arising from our finite $n_{\rm samp} = 10^6$ (Sec.\ \ref{subsec:IIB}).}
\label{fig:spatiallyaveragedmetrics}
\end{figure*}

Next, to relate the trends discussed above to other trends that will be relevant for understanding these packings, we examine three additional spatially-averaged structural-order metrics.
Figure \ref{fig:spatiallyaveragedmetrics} shows the variation of $\langle|\Psi_6(i)|\rangle$, $S_{\rm loc}(100)$, and  $\langle \delta\phi(100) \rangle$ with $\alpha$ for selected $\lambda$.
The variation of these quantities with $\lambda$ for selected $\alpha$ is discussed in the Appendix.

The results shown in panel (a) can be most easily understood in terms of the ``excess'' local hexatic order $\mathcal{O}_{\rm hex} \equiv \langle|\Psi_6(i)|\rangle - \langle|\Psi_6|\rangle_{\rm rand}$, where $\langle|\Psi_6|\rangle_{\rm rand} \simeq 0.366$ is the mean value for a particle with six randomly-oriented neighbors.
Negative $\mathcal{O}_{\rm hex}(\alpha)$ imply that six nearest neighbors lying  at $60^\circ$ angles with respect to one another is less likely in the corresponding jammed packing than it is in a random medium.
All $\alpha > 2$ packings have negligible or negative $\mathcal{O}_{\rm hex}$.
This result was expected because increasing $\alpha$ makes it more difficult for a particle's six nearest neighbors to lie at $60^\circ$ angles with respect to one another, as is required to achieve  $|\Psi_6(i)| = 1$ [Eq.\ \ref{eq:Psi6ofi}]. 
$\mathcal{O}_{\rm hex}(\alpha)$ also decreases rapidly with increasing $\lambda$ for $\alpha < 2$ and $\lambda > 1.4$.
One reason for this trend is that the increasing $\Delta Z(\lambda)$ [Fig.\ \ref{fig:phiandZ1}] make summing over the \textit{six} nearest neighbors of particle $i$ [Eq.\ \ref{eq:Psi6ofi}] less-well-justified.

For $\lambda \leq 1.4$, systems are able to form nearly-perfect local hexatic order when $\alpha - 1 \ll 1$, with values of $\langle|\Psi_6(i)|\rangle$ as high as $0.976$.
This trend is expected for slow LS-like particle growth.\cite{lubachevsky91}
We find that $\langle|\Psi_6(i)|\rangle$ actually increases with $\alpha$ over the range $1 \leq \alpha \leq 1.05$, perhaps because the rapidly increasing $Z_{\rm J}(\alpha)$ [Fig.\ \ref{fig:phiandZ1}(b)] makes it more likely that the six nearest neighbors contibuting to $\Psi_6(i)$ [Eq.\ \ref{eq:Psi6ofi}] are all contacts. 
Over the range $1 \leq \alpha \leq 2$, these systems' $\mathcal{O}_{\rm hex}(\alpha)$ are approximately described by the Gaussian functional form
\begin{equation}
\mathcal{O}_{\rm hex}(\alpha) = A\exp\left[-\frac{(\alpha-\alpha_0)^2}{2a^2}\right],
\label{eq:GaussianPsi6}
\end{equation}
where $\alpha_0 \simeq 1.05$ and ($A,a$) are $\lambda$-dependent constants.

The affinely-stretched hexagonal lattices shown in Fig.\ \ref{fig:xtal} also exhibit a Gaussian-like variation of $\langle|\Psi_6(i)|\rangle$ with $\alpha$, with comparable ($A,a$) but $\alpha_0 = 1$. 
We find that the jammed packings for $\lambda \lesssim 1.4$ actually have \textit{higher} local hexatic order than these crystals over a wide range of $\alpha$.
In other words, particles' nearest neighbors are more likely to lie at approximately-$60^\circ$ angles with respect to each other in the jammed packings than the corresponding monodisperse crystals.
This might illustrate the limitations of $\langle|\Psi_6(i)|\rangle$ as a structural-order metric for systems of anistropic particles; for example, monodisperse ellipse crystals with $\alpha \simeq 2.064178$ have $\langle|\Psi_6(i)|\rangle = 0$.
On the other hand, it also suggests that the jammed packings with $\alpha \simeq \alpha_{\rm max}$ and ``smallish'' $\lambda$ are able to maintain a high level of hexatic order by forming motifs that are not found in the corresponding crystals.
We will investigate this possibility further in Section \ref{subsec:IIIC}.

The results shown in panel (b) can be most easily understood in terms of the competition between maximizing the jammed packings' density and minimizing their free energies during the particle-growth process.
For monodisperse ellipses, the equilibrium nematic-liquid-crystalline phase appears only for  $\alpha > \alpha^{\rm nem}_{\rm mono} \simeq 2.4$.\cite{bautista14b}
While the equilibrium solid-nematic-crystalline-phase exists for all $\alpha > 1$ in the limit $\phi \to \phi_{\rm xtal}$, densifying systems with $\alpha \lesssim 1.6$ must pass through a range of $\phi$ where the equilibrium solid phase is a \textit{plastic} (positionally-ordered but orientationally-disordered) crystal to reach this point.\cite{bautista14b}
Moreover, the structure of  their saturated random-sequential-adsorption packings is  tip-to-side-contact-dominated for $\alpha \lesssim \alpha^{\rm nem}_{\rm mono}$;\cite{abritta22} this should further disfavor formation of locally-nematic structure as systems with $\alpha \lesssim \alpha^{\rm nem}_{\rm mono}$  are passing through the $\phi \simeq \phi_{\rm RSA}(\alpha)$ regime.\footnote[3]{$\phi_{\rm RSA}(\alpha)$ lies within the range $0.498 \lesssim \phi_{\rm RSA} \lesssim 0.583$ for the $\lambda = 1$ systems considered here.\cite{abritta22}}
Thus, for $\alpha <  \alpha^{\rm nem}_{\rm mono}$ (and especially for $\alpha \lesssim 1.6$), several factors mitigate against formation of jammed packings with intermediate- or long-range nematic order, even though this ordering allows $\phi_{\rm J}$ to be maximized in the $\lambda \to 1$ limit.

\begin{figure*}
\centering
\includegraphics[width=7in]{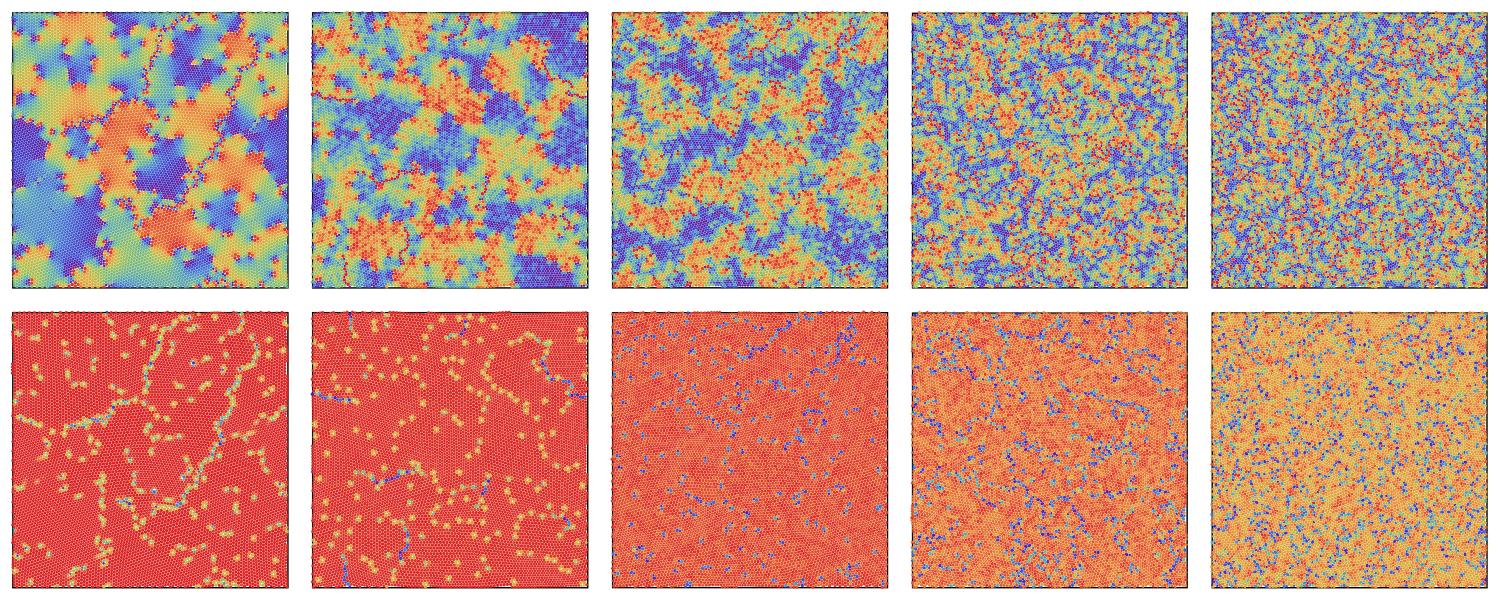}
\vspace{-10pt}
\caption{Evolution of fractionation and hexatic order with increasing $\alpha$ for $\lambda = 1.4$.  Columns have $\alpha = 1, 1.1, 1.2, 1.3, 1.4$ from left to right.  In the top (bottom) row, particle colors range from purple to red as $\sigma_i$ increases from $\sigma_{\rm min}$ to $\sigma_{\rm max}$ (as $|\Psi_6(i)|$ varies from $0$ to $1$).}
\label{fig:snaps1}
\end{figure*}

\begin{figure*}
\centering
\includegraphics[width=7in]{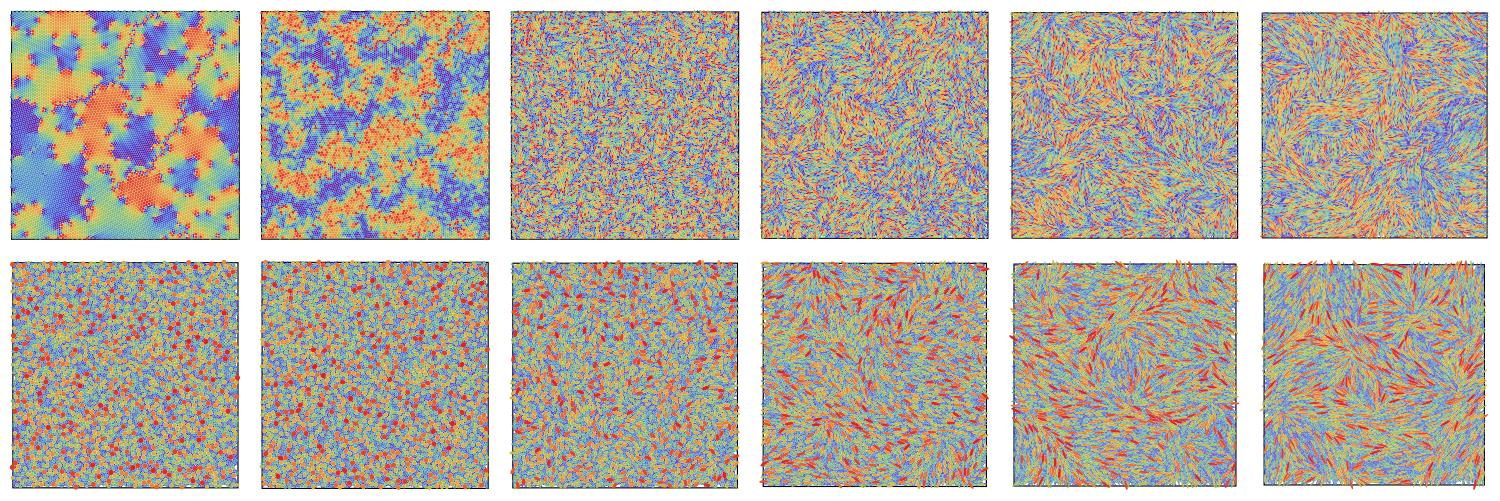}
\vspace{-10pt}
\caption{Evolution of fractionation with both $\alpha$ and $\lambda$.  Columns have $\alpha = 1, 1.2, 2, 3, 4, 5$ from left to right, while the top and bottom rows respectively have $\lambda = 1.4$ and $\lambda = 5$.   Particle colors range from purple to red as $\sigma_i$ increases from $\sigma_{\rm min}$ to $\sigma_{\rm max}$.}
\label{fig:snaps2}
\end{figure*}

Our results for $S_{\rm loc}(100)$ are largely consistent with the above hypotheses.
With the exception of $(\alpha,\lambda) = (1.025,1)$, none of the jammed states with $\alpha \leq 1.6$ possess significant intermediate-range nematic ordering.
The $\partial S/\partial \alpha$ increase rapidly for $\alpha \gtrsim 1.6$, and reach peaks at $\alpha \simeq  \alpha^{\rm nem}_{\rm mono}$ before decreasing again for $\alpha \gtrsim 3$.
One surprising feature is that the degree of nematic order formed during particle growth for systems with $\alpha >  \alpha^{\rm nem}_{\rm mono}$ increases rapidly over the range $1 \leq \lambda \lesssim 2$, and saturates only for $\lambda \gtrsim 3$.
Moreover, our $\lambda = 2$ systems with $\alpha \gtrsim 2$ show notably-stronger intermediate-range nematic order than those examined in a very recent study\cite{jiang26} of $\lambda = 2$ systems that employed fast particle growth.
These results indicate that SWAP becomes more effective at generating nematic domains including $\gtrsim 100$ particles as $\lambda$ increases.
We will further explore their implications in Section \ref{subsec:IIIC}.

Several features of the average intermediate-range packing-fraction fluctuations $\delta\phi(100)$ are apparent from panel (c).
First, they decrease \textit{substantially} over the range $1 \leq \lambda \leq 1.4$ for all $\alpha$.
The associated negative $(\partial \delta\phi/\partial \lambda)_\alpha$ arise primarily from SWAP's efficacy in promoting maximally-efficient ellipse packing;\cite{hoy24b} here we find that they are enhanced further by employing slow particle growth.
Second, the decreases reverse [i.e., $(\partial \delta\phi/\partial \lambda)_\alpha$ becomes positive] as $\lambda$ increases further, but they do so in a strongly-$\alpha$-dependent fashion.
Since both $S_{\rm loc}(100)$ and $\delta\phi(100)$ are nearly $\lambda$-independent for $\alpha \gtrsim 4$ and $\lambda \gtrsim 3$, we expect that the later reversals for larger $\alpha$ are directly associated with the increasing nematic order.

Given our focus on $\phi > \phi_{\rm xtal}$ packings, perhaps the most noteworthy aspect of Fig.\ \ref{fig:spatiallyaveragedmetrics} is that none of these spatially-averaged metrics exhibit dramatic changes as systems cross the $\lambda_{\rm xtal}(\alpha)$ boundary shown in Fig.\ \ref{fig:xtal}.
Next we will show that looking at how the fluctuations in these metrics (about their average values) are \textit{spatially correlated} tells a much-more-interesting story.

\subsection{Spatially-correlated structural order}
\label{subsec:IIIC}

Figure \ref{fig:snaps1} shows that fractionation and local hexatic order are intimately connected in packings with small ($\alpha,\lambda$). 
Disks with $\lambda = 1.4$ form high-quality fractionated polycrystals in which the grains consist primarily of intermediate-size particles and the grain boundaries include the largest and smallest particles.\cite{hoy24b}
As expected, local hexatic order remains very high within the grains but drops dramatically at the grain boundaries.
The spatially-extended nature of these boundaries clearly indicates long-range correlations in $|\Psi_6(i)| - \langle |\Psi_6(i)| \rangle$.
Moreover, voids are clearly visible within them.

\begin{figure*}
\centering
\includegraphics[width=7in]{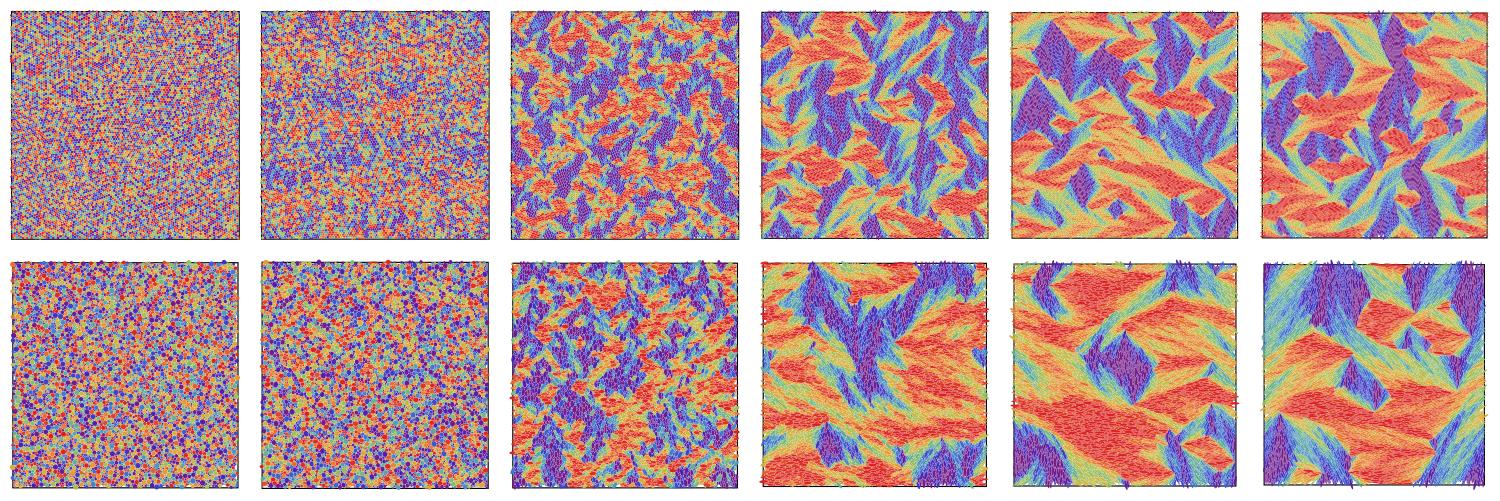}
\vspace{-10pt}
\caption{Evolution of nematic order with both $\alpha$ and $\lambda$.  Columns have $\alpha = 1, 1.2, 2, 3, 4, 5$ from left to right, while the top and bottom rows respectively have $\lambda = 1.4$ and $\lambda = 5$.  Particle colors range from purple to red as $\cos^2(\theta_i)$ varies from $0$ to $1$.}
\label{fig:snaps3}
\end{figure*}

\begin{figure*}
\centering
\includegraphics[width=7in]{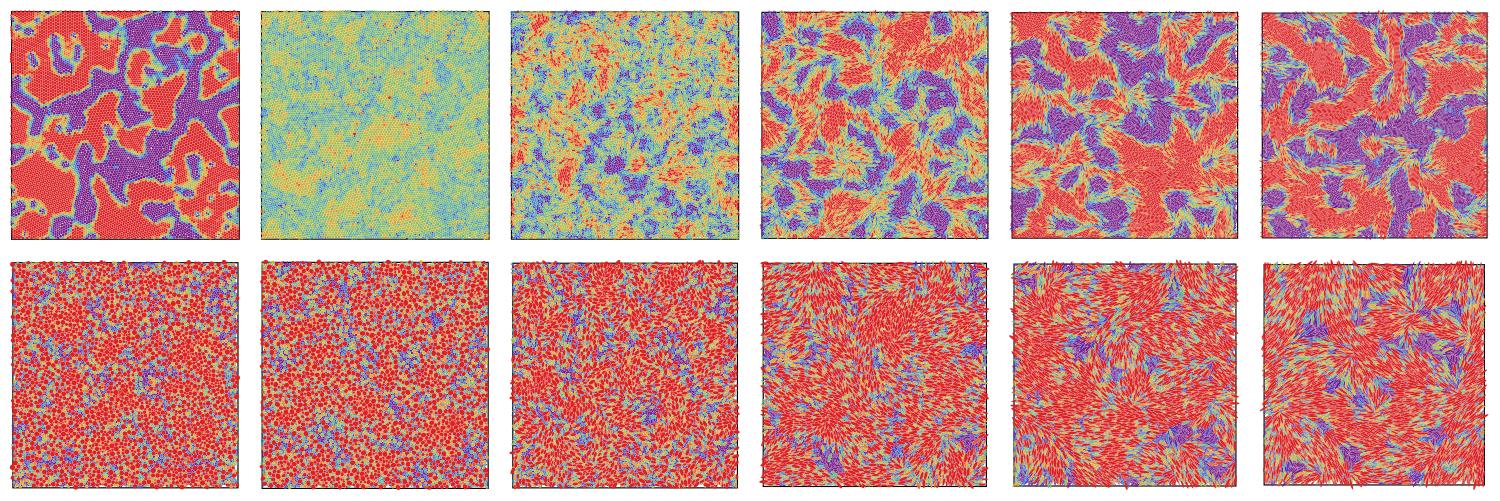}
\vspace{-10pt}
\caption{Evolution of packing-fraction fluctuations with both $\alpha$ and $\lambda$.  Columns have $\alpha = 1, 1.2, 2, 3, 4, 5$ from left to right, while the top and bottom rows respectively have $\lambda = 1.4$ and $\lambda = 5$.  Particle colors range from purple to red as $\phi_i(100) - \langle \phi_i(100) \rangle$ varies from 
$-5\times 10^{-3}$ (or less) to $-5\times 10^{-3}$ (or more).}
\label{fig:snaps4}
\end{figure*}

As expected from the slower particle growth and larger system size (i.e., $N = 10^4$ rather than $N = 10^3$) employed in the present study, this fractionated-polycrystalline order extends to larger $\alpha$ values than in Ref.\ \cite{hoy24b}.
The overall trends, however, are similar.
As $\alpha$ increases, fractionation weakens gradually for $\alpha \lesssim \alpha_{\rm max}$ and then rapidly for  $\alpha \gtrsim \alpha_{\rm max}$.
While $\langle |\Psi_6(i)| \rangle$ decreases relatively slowly over the range $1 \leq \alpha \lesssim \alpha_{\rm max}$ [as shown in Fig.\ \ref{fig:spatiallyaveragedmetrics}(a)], the spatial heterogeneity of this order decreases \textit{rapidly} over the same range.
Systems with $\alpha = \alpha_{\rm max} = 1.2$ retain a large $\langle |\Psi_6(i)| \rangle$, but lack the the clear grain boundaries visible for $\alpha = 1$.
Instead, the particles with low $|\Psi_6(i)|$ values are isolated.
The spatial heterogeneity of $|\Psi_6(i)|$ then increases slowly with $\alpha$ for $\alpha \gtrsim \alpha_{\rm max}$. 
Thus systems with $\alpha \simeq \alpha_{\rm max}$ and smallish $\lambda$ appear to combine \textit{homogeneous} local hexatic order with  $\langle |\Psi_6(i)| \rangle$ that are comparable to or greater than to those found the monodisperse crystals.

Figure \ref{fig:snaps2} shows how fractionation varies over a wider range of $\alpha$ and $\lambda$.
For $\lambda = 1.4$, it remains negligible over a broad range ($1.5 \lesssim \alpha \lesssim 4$)  before increasing again for $\alpha \gtrsim 4$.
For $\lambda = 5$, we find \textit{microfractionation} of the type discussed above, i.e.\ a higher propensity for the largest particles to be surrounded by many small neighbors and the smallest particles to be trapped between three large neighbors, for all $\alpha$.
Macro-scale fractionation (as indicated by long-range correlations in $\sigma_i - \langle \sigma_i \rangle$), however, remains minimal in these systems for all $\alpha \lesssim 4$.
In particular, they do \textit{not} form fractionated plastic crystals.

Figure \ref{fig:snaps3} illustrates how nematic order remains minimal over the range $1 < \alpha \lesssim 1.6$, but increases rapidly with $\alpha$ for $\alpha \gtrsim 2$.
Systems with smaller aspect ratios (including the two leftmost systems in the bottom row) can develop a denser-than-crystalline structure without developing even intermediate-range nematic order.
For larger $\alpha$, the size of the nematic grains for fixed $\alpha$ increases substantially with $\lambda$ over the range $1 \leq \lambda \lesssim 3$ before plateauing for $3 \lesssim \lambda \leq 5$.
The latter trend shows that the saturation in $S_{\rm loc}(100)$ illustrated in Fig.\ \ref{fig:spatiallyaveragedmetrics}(b) is associated with the size rather than the ``intensity'' of the nematic grains.
The former trend is consistent with the positive $\partial \phi_{\rm J}/\partial \lambda$, and also with previous work showing that increasing polydispersity shifts the onset of isotropic-nematic coexistence (in equilibrium systems) to lower $\phi$.\cite{lekkerkerker84}
Another possibility is that alignment of larger particles drives alignment of nearby smaller particles as $\phi \to \phi_{\rm J}$ during particle growth.
It would be very interesting to investigate the development of anisotropically-correlated local nematic order using the sophisticated cluster-based measures employed in Ref.\ \cite{jiang26}, but doing so is beyond our present scope.

Figure \ref{fig:snaps4} shows that the spatial heterogeneity of $\langle \phi_i(\mathcal{N}) \rangle$ also varies strongly and nontrivially with both $\alpha$ and $\lambda$.
For $\lambda = 1.4$, disk packings'  $\langle \phi_i(\mathcal{N}) \rangle$ exhibit very strong long-ranged fluctuations; their  $\langle \phi_i(\mathcal{N}) \rangle$ appear to be largest within the highest-quality hex-crystalline grains and lowest near the grain boundaries.
As $\alpha$ increases, spatial correlations of fluctuations in their $\langle \phi_i(\mathcal{N}) \rangle$ weaken rapidly over the range $1 < \alpha \lesssim \alpha_{\rm max}$.
Notably, these images suggest that (as has been previously reported for MRJ disk and ellipse packings\cite{zachary11,zachary11b,zachary11c}) minimal correlation of  $\langle \phi_i(\mathcal{N}) \rangle$ can coexist with strong fractionation.
These correlations remain weak until $\alpha$ increases into the range where intermediate-range nematic order develops, but thereafter strengthen rapidly and are clearly visible for $\alpha \geq 2$.

For $\lambda = 5$ and very small $\alpha$, the spatial correlations of fluctuations in $\langle \phi_i(\mathcal{N}) \rangle$ are substantially weaker than they were for $\lambda = 1.4$, presumably because the hex-crystalline grains are absent.
In contrast to the results for $\lambda = 1.4$, these correlations do not appear to weaken significantly as $\alpha$ increases towards $\alpha_{\rm max}$.
Instead, their strength and range appears to remain nearly constant until until $\alpha$ increases into the range where intermediate-range nematic order develops.
Finally, although they indeed strengthen with increasing $\alpha \gtrsim 3$, they do so less dramatically than their $\lambda = 1.4$ counterparts.

Overall, the results presented in Figs.\ \ref{fig:snaps1}-\ref{fig:snaps4} indicate that macro-scale fractionation in ellipse packings (i) produces a high degree of either local hexatic or local nematic order, and (ii) promotes strong packing-fraction fluctuations over comparable length scales \textit{except when} $\alpha \simeq \alpha_{\rm max}$ and $\lambda$ is ``smallish''.
In the following section, we will focus on this special case.

\subsection{Hyperuniformity, triangulation, and potential connections to ideal glasses}
\label{subsec:IIID}

Hyperuniform systems have numerous desirable properties.\cite{torquato18b}
Early work showed that packings of monodisperse disks and spheres that are (i) maximally-random, (ii) saturated (i.e.\ have no voids into which another particle can be placed without producing overlaps), and (iii) strictly jammed are always (iv) hyperuniform with respect to number-density fluctuations, i.e.\ their structure factors $S(k)$ have $S(k) \sim k^a$ as $k \to 0$, with $a > 0$.\cite{torquato00,torquato01,torquato03,torquato18b}
More recent studies\cite{zachary11,zachary11b,zachary11c} showed that packings of polydisperse disks, superdisks and ellipses with properties (i-iii) lack property (iv) owing to fractionation, but can remain hyperuniform with respect to \textit{packing-fraction fluctuations}, i.e.\ their spectral densities $\hat{\chi}_{\rm V}(k)$ have $\hat{\chi}_{\rm V}(k) \sim k^a$ as $k \to 0$, with $a > 0$.

\begin{figure}
\centering
\includegraphics[width=3.375in]{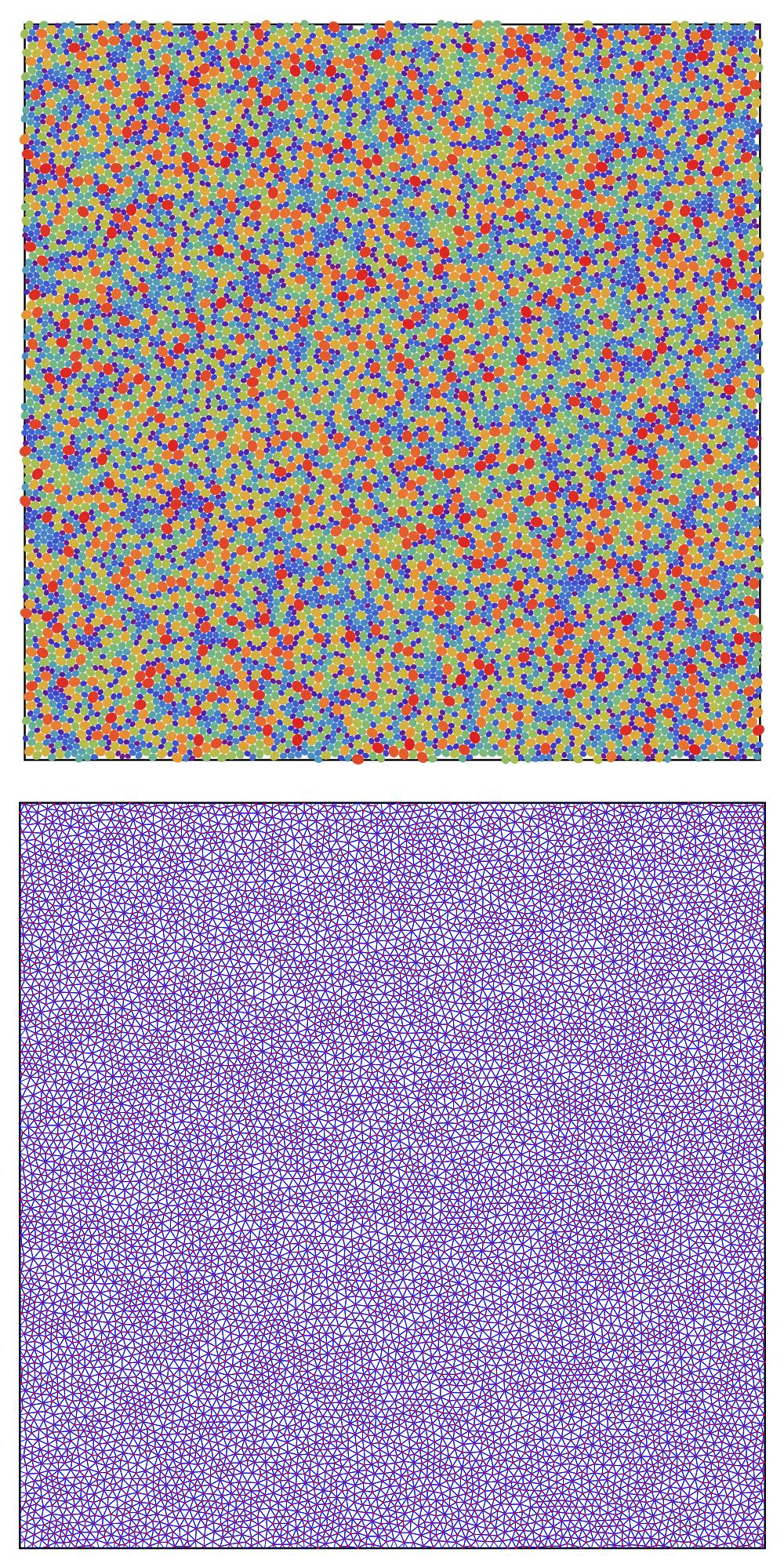}
\caption{This $\alpha = 1.2$ $\lambda = 2.1$ packing has $\phi_{\rm J} = 0.9087$, $Z_{\rm J} = 6$ (exactly), and is triangulated.  Colors in the top panel range from purple to red as $\sigma_i$ increases from $\sigma_{\rm min}$ to $\sigma_{\rm max}$.}
\label{fig:triang1}
\end{figure}

Ref.\ \cite{bolton26} showed that ideal-glass-like disk packings are hyperuniform with respect to packing-fraction fluctuations and also \textit{triangulated}.
The latter property implies that the ``faces'' of their contact graphs consist entirely of triangles.
This occurs when all particles' first coordination shells are closed, i.e.\ contain no gaps.
Since the number of edges in these graphs is $E = ZN/2$, each triangle has 3 edges, and each edge is shared by two triangles, the total number of faces in these graphs is $F = ZN/3$.
When $Z = 6$, $F = 2N$, and since each face corresponds to one void, such systems have exactly $2N$ disconnected voids.
As a consequence, their void-space geometry is highly constrained.
Refs.\ \cite{zachary11,zachary11b,zachary11c} argued that highly-constrained void-space geometry 
promotes packing-fraction hyperuniformity in strictly-jammed MRJ-like packings.\cite{torquato00,torquato01}
They also suggested that this type of hyperuniformity results from the competing requirements of strict and maximally-random jamming.

Most of the systems considered here have much-higher $\phi_{\rm J}$ and $Z_{\rm J}$ compared to the jammed packings studied in Refs.\  \cite{delaney05,donev07,vanderwerf18,rocks23,hoy24b,jiang26}, and many of them are substantially-locally-ordered according to metrics such as $\langle|\Psi_6(i)|\rangle$ and $S_{\rm loc}(100)$.
These systems are therefore  far from the minimally-ordered ``MRJ'' limit.\cite{torquato00}
Moreover, as illustrated in Fig.\ \ref{fig:snaps4}, many of them clearly have large long-wavelength packing-fraction fluctuations and therefore cannot be hyperuniform.
Taken together, these results raise the question of whether our algorithm generates ideal-glass-like (i.e., triangulated and hyperuniform) ellipse packings for \textit{any} combinations of ($\alpha,\lambda$).

Here we show that it  \textit{may}.
For $\alpha = 1.2$, single runs of the algorithm produced exactly-isostatic, triangulated packings for $\lambda = 2.1$, $2.2$, $2.4$ and $2.8$.
Figure \ref{fig:triang1} shows the configuration and contact graph for $\lambda = 2.1$.
The triangulation is nontrivial.
Only 42\% of the particles have $Z_i = 6$, and -- in contrast to the $\lambda = 1.4$ packing (Fig.\ \ref{fig:ZiVsSigmaI}) -- there are no long-range contact-number correlations.
This packing also lacks any other obvious long-range spatially-correlated fluctuations in its structural order (as identified by the metrics we have investigated above) such as fractionation, hexatic grain boundaries, or nematically-ordered domains.

\begin{figure}[htbp]
\centering
\includegraphics[width=3in]{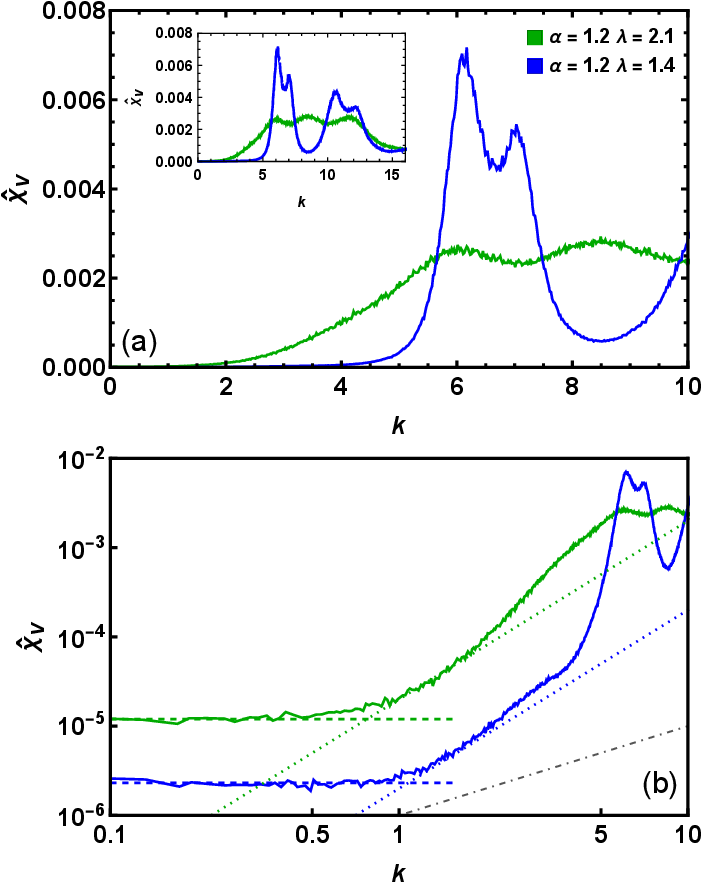}
\caption{Spectral densities for two of the $(\alpha,\lambda)$ combinations highlighted above.
All results are averaged over 25 independently-prepared samples, and all lengths are scaled so that $\langle \sigma \rangle = 1$.
The inset to panel (a) shows the same data plotted over a wider range of $k$.
In panel (b), dashed lines indicate our estimated $\hat{\chi}_{\rm V}(0)$.
Dotted and dot-dashed lines respectively illustrate $\hat{\chi}_{\rm V}(k) \sim k^2$ and $\hat{\chi}_{\rm V}(k) \sim k$ scaling, and are intended to serve as guides to the eye.}
\label{fig:ChiOfK}
\end{figure}

Figure \ref{fig:ChiOfK} contrasts the ensemble-averaged $\hat{\chi}_{\rm V}(k)$ of an $(\alpha,\lambda)$ combination producing triangulation with those of the $(\alpha,\lambda)$ combination that minimizes $\delta \phi(100)$ and $\Delta Z_{\rm J}$.
Visual inspection panel (a) indicates multiple difference from the bidisperse $(\alpha,\lambda) = (1.1,1.4)$ and $(1.4,1.4)$ packings studied in Refs.\ \cite{zachary11,zachary11c}.
First, our packings' $\hat{\chi}_{\rm V}(k)$ appears to be effectively zero in the small-$k$ limit rather than exhibiting $\hat{\chi}(k) \sim k$ (i.e.\ $a = 1$) scaling in this limit.
Second, for slightly larger $k$, our packings' $\hat{\chi}_{\rm V}(k)$ are concave up rather than concave down.
Third, our packings' $\hat{\chi}_{\rm V}(k)$ exhibit three or four peaks over the range $0 \leq k\langle\sigma\rangle \leq 15$ rather than two.
We also find that -- as expected -- increasing $\lambda$ broadens the peaks in $\hat{\chi}_{\rm V}(k)$.

Plotting the same results on log scale [panel (b)] yields additional insights.
First, both $(\alpha,\lambda)$ combinations' $\hat{\chi}_{\rm V}(k)$ are nearly constant for $k \ll 1$, i.e.\ they both have $a = 0$ in the small-$k$ limit.
We find $\hat{\chi}_{\rm V}(0) \simeq 2.3\times 10^{-6}$ for $(\alpha,\lambda) = (1.2,1.4)$ and $1.2 \times 10^{-5}$ for  $(\alpha,\lambda) = (1.2,2.1)$.
Second, we see no evidence of a $\hat{\chi}(k) \sim k$ scaling regime.
Instead, both $(\alpha,\lambda)$ combinations cross over into a narrow $\hat{\chi}(k) \sim k^2$ scaling regime at $k \simeq 1$.
Third, their  $\hat{\chi}_{\rm V}(1)$ are much smaller than those reported in Refs.\ \cite{zachary11,zachary11c}.
All three trends presumably arise from the combination of slow particle growth and SWAP moves promoting a more uniform distribution of void sizes; this is to be expected in packings with larger $\phi_{\rm J}$.

The question of whether these packings are ``truly'' hyperuniform remains unresolved.
The $\lim_{k \to 0} \hat{\chi}(k) \sim k^0$ scalings suggest that they are not, while the extremely-low $\hat{\chi}_{\rm V}(0)$ values suggest that they are.
Ref.\ \cite{chen18} suggested that systems with $H = \hat{\chi}_{\rm V}(0)/\hat{\chi}_{\rm V}(k = k_{\rm max}) \leq 10^{-2}$,  where $\hat{\chi}_{\rm V}(k = k_{\rm max})$ is the spectral density's maximum magnitude, are \textit{effectively} hyperuniform.
Our packings have $H(1.2,1.4) \simeq 3\times 10^{-4}$ and $H(1.2,2.1) \simeq 4\times 10^{-3}$, so they satisfy this criterion.
Ref.\ \cite{maher24} showed that rigorously establishing a packing's hyperuniformity requires it to be strictly jammed.\cite{torquato01}
Although the isostatic (or near-isostatic), rattler-free character of the packings discussed in this section suggests that they are close to satisfying this criterion, their $Z_i$ values were calculated using a fairly loose contact tolerance $C = 10^{-3}$ (Section \ref{subsec:IIB}), and we have not tested them for strict jamming.
Modifying our packing-generation algorithm to achieve strict jamming (as defined in Section VC of Ref.\ \cite{donev07}) could lead to both stronger hyperuniformity and a clearer resolution of ultrastable disordered ellipse packings' optimal $\hat{\chi}_{\rm V}(k) \sim k^a$ scaling in the limit $k \to 0$.

\subsection{Lowest-dispersity $\phi > \phi_{\rm xtal}$ vs.\ densest packings}
\label{subsec:IIIE}

A skeptical reader might claim that demonstrating the existence of $\phi > \phi_{\rm xtal}$ ellipse packings is rather trivial because one can access the limit $\phi \to 1$ simply by preparing RAPs using a method such as the one described in Ref.\ \cite{delaney08}.
On the other hand, such RAPs would not exhibit the complex multi-scale order discussed above.
We will now highlight some additional characteristic features of these packings' structural order.

\begin{figure}
\centering
\includegraphics[width=3.375in]{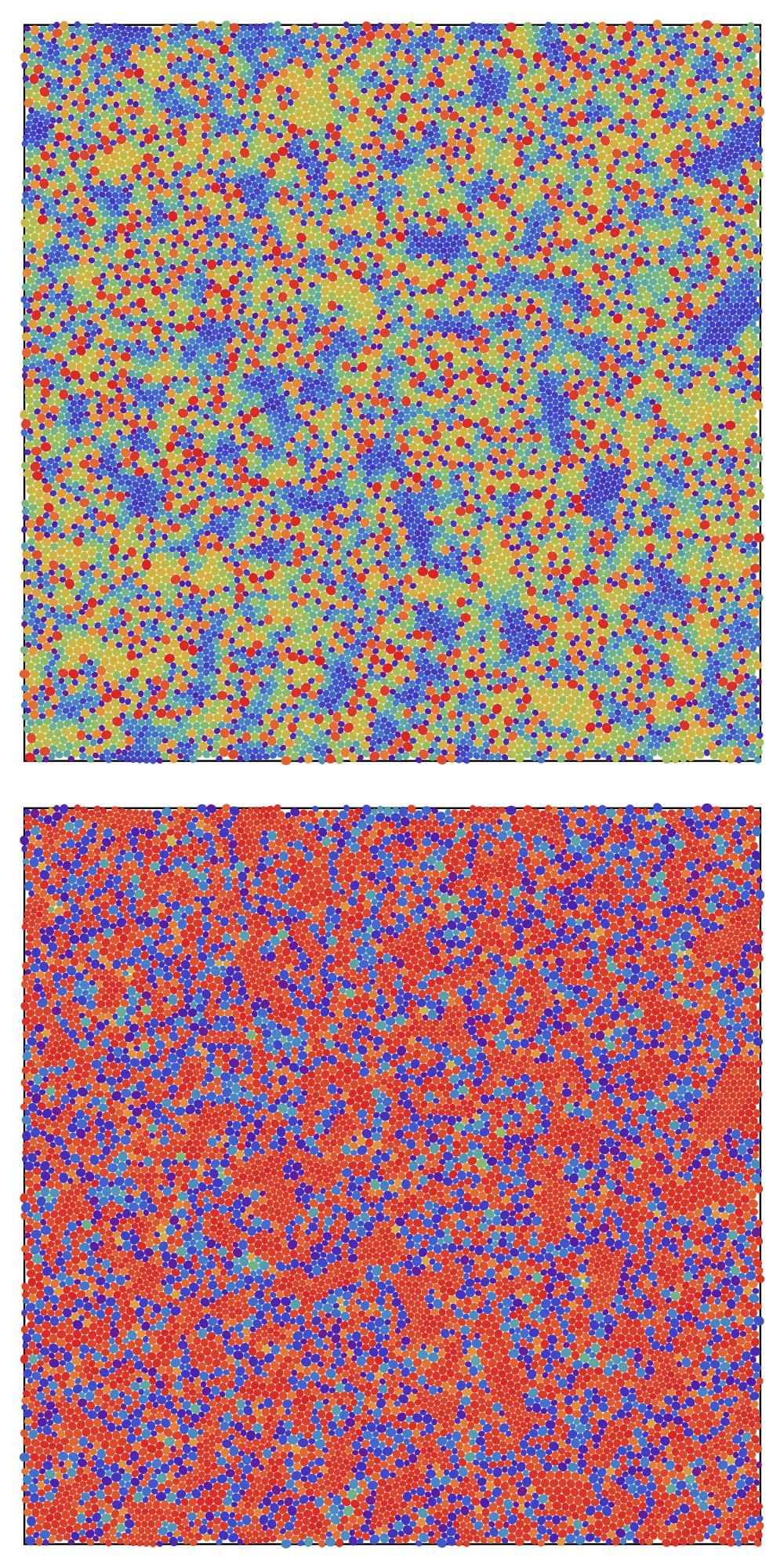}
\caption{The least-disperse $\phi > \phi_{\rm xtal}$ ellipse packing we found has $\alpha = 1.15$, $\lambda = 1.65$, $\phi_{\rm J} = 0.9070$ and $Z_{\rm J} = 5.9948$.  Colors in the top panel range from purple to red in order of increasing $\sigma_i$, while those in the bottom panel range from purple to red as $|\Psi_6(i)|$ varies from $0$ to $1$.}
\label{fig:snaps6}
\end{figure}

The lowest-dispersity $\phi > \phi_{\rm xtal}$ ellipse packing we found has $(\alpha,\lambda) = (1.15,1.65)$ and is shown in Figure \ref{fig:snaps6}.\footnote[4]{The  $\lambda_{\rm xtal}^{\rm min} \simeq 1.63$ value quoted above was estimated by assuming that $\phi_{\rm J}(1.15,\lambda)$ is linear over the range $1.60 \leq \lambda \leq 1.65$.}
This packing might be best described as a ``fractionated plastic hexatic polycrystal''.
While its $\langle | \Psi_6(i) | \rangle = 0.655$ is significantly below the value for  the corresponding monodisperse lattice ($\langle | \Psi_6(i) | \rangle = 0.954$), almost one-third of its constituent particles have $| \Psi_6(i) | >  0.954$.
The individual hexatic-crystalline grains consist mostly of intermediate-sized particles.
Focusing on the small purple dots (large red dots) reveals that the smallest (largest) particles tend to appear at isolated positions within grains formed primarily by much larger particles (at the boundaries between different grains).

\begin{figure}
\centering
\includegraphics[width=3.375in]{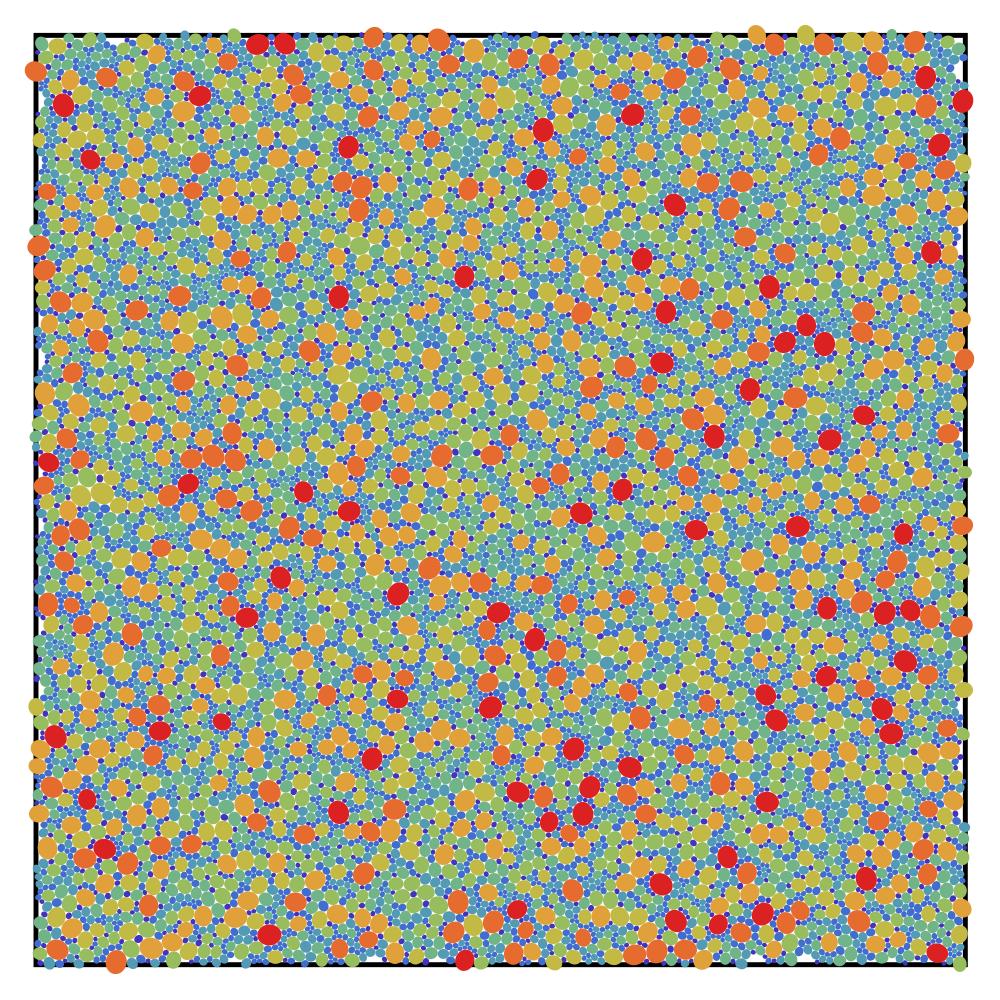}
\caption{The densest  packing we found has $\alpha = 1.15$, $\lambda = 5$, $\phi_{\rm J} = 0.9192$ and $Z_{\rm J} = 5.998$.  Colors range from purple to red as $Z_i$ increases from $3$ to $\geq 12$.}
\label{fig:snaps7}
\end{figure}

As illustrated in Figure \ref{fig:snaps7}, the structure of the densest packings produced in this study is radically different.
This $(\alpha,\lambda) = (1.15,5)$ packing has two characteristic motifs: (i) the largest particles are often surrounded by at least twelve small neighbors, and (ii) the smallest particles are often trapped between three large neighbors.
These motifs are also present in the abovementioned $\lambda = \lambda_{\rm crit}$ bidisperse crystals (with ``often'' replaced by ``always'' and ``at least twelve'' replaced by ``exactly twelve''\cite{uche04}), 
but here they appear within packings that lack any obvious long-range structural order.
The same motifs also appear regularly in other high-$\phi_{\rm J}$ disordered packings with larger $\alpha$ and $\lambda \gtrsim 4$.

\section{Discussion and Conclusions}
\label{sec:conclude}

In this paper, we have shown that integrating SWAP Monte Carlo moves into a slow LS-like particle growth algorithm allows one to generate disordered jammed packings of modestly-polydisperse ellipses  that are denser than the corresponding infinitely-fractionated locally-monodisperse crystals for a wide range of $\alpha$.
While we cannot claim that these packings are the thermodynamic ground states for the given $(\alpha,\beta,\lambda)$  -- denser infinitely-fractionated locally-bidisperse crystals may exist\cite{toth64} -- they nonetheless are the first reported disordered $\phi > \phi_{\rm xtal}^{\rm mono}$ packings of \textit{anisotropic} particles.

We also performed a wide range of structural analyses which collectively indicate that while these packings' characteristic motifs are strongly $\alpha$- and $\lambda$-dependent, none of the investigated metrics exhibit sharp changes as $\lambda$ exceeds the minimal size ratio $\lambda_{\rm xtal}(\alpha)$ defined by $\phi_{\rm J}(\alpha,\lambda) \geq  \phi_{\rm xtal}^{\rm mono}$.
In other words, there does not appear to be an \textit{athermal} phase transition associated with the emergence of denser-than-crystalline disordered packings.
This was not a priori obvious, given that there must be a corresponding \textit{thermal} phase transition associated with the exchange of stability (away from the abovementioned infinitely-fractionated locally-monodisperse crystals) in the $P \to \infty$ limit.

$\lambda_{\rm xtal}(\alpha)$ exhibits an intriguing non-monotonic variation with $\alpha$.
First it drops sharply, from $2.17$ for $\alpha = 1$ to a minimum value $\lambda_{\rm xtal}^{\rm min} \simeq 1.63$ for $\alpha \simeq 1.15$.
This drop naturally mirrors the rapid increases in $\phi_{\rm J}(\alpha,\lambda)$ over the same range of $\alpha$, but is worth interpreting further here.
Our results suggest that polydisperse ellipses'  $\phi_{\rm J}(\alpha)$ is maximized and their $\lambda_{\rm xtal}(\alpha)$ is minimized at $\alpha \simeq \alpha_{\rm max} = 1.2$ in the slow-particle-growth limit because this value of $\alpha$ optimizes particles' ability to form denser packings by rotating away from each other\cite{donev04,donev07} while retaining their ability to form packings with high local hexatic order.

As $\alpha$ increases beyond $\alpha_{\rm max}$,  $\lambda_{\rm xtal}$ increases steadily  until it reaches a local maximum ($\sim 2.46$) at $\alpha \simeq 2.3$.
In addition to mirroring the decreases in $\phi_{\rm J}(\alpha,\lambda)$ over the same range of $\alpha$, this increase reflects the fact that packings' local hexatic order diminishes rapidly while their intermediate-scale nematic order remains small over most of this range.
When both types of order remain limited, so is the ability of particles to form the  locally-efficient motifs that are an essential feature of disordered $\phi > \phi_{\rm xtal}$ packings.
We believe that the maximum in $\lambda_{\rm xtal}(\alpha)$ at $\alpha \simeq 2.3$ is associated with the peak in $\partial S_{\rm loc}(100)/\partial \alpha$ (Fig.\ \ref{fig:spatiallyaveragedmetrics}); this peak corresponds to the emergence of an equilibrium nematic-liquid-crystalline phase.\cite{bautista14b}
It is remarkable, however, that this emergence of intermediate-range nematic order has only limited effects on $\lambda_{\rm xtal}(\alpha)$.

As $\alpha$ increases further, $\lambda_{\rm xtal}$ passes through a minimum at $\alpha \simeq 3$ and thereafter increases slowly but steadily, remaining below $2.88$ all the way out to $\alpha = 5$.
This trend mirrors the slowly decreasing $\phi_{\rm J}$, and also appears to be associated with the decreasing $Z_{\rm J}$ and increasing $\delta\phi(100)$, over the range $3 \lesssim 5$.
All four of these trends can be explained by increasing $\alpha$ making it more difficult for systems with fixed $\lambda$ to efficiently pack the boundaries between different nematic domains.
In other words, the ability of systems with broad $P(\sigma)$ distributions to form disordered packings that are homogeneously-dense and nearly-isostatic despite their broad $P(Z_i)$ distributions decreases with increasing $\alpha \gtrsim 3$.

Our lowest values of $\lambda_{\rm xtal}(\alpha)$ are substantially smaller than might have been expected for a ``physical'' ellipse-packing-generation protocol.
While Toth showed long ago that bidisperse disks can form ordered $\phi > \phi_{\rm xtal}^{\rm mono}$ packings for all $\lambda \gtrsim 1.5$,\cite{toth64}  disordered $\phi > \phi_{\rm xtal}^{\rm mono}$ packings of continuously-polydisperse disks with comparably-small $\lambda$ values were not reported until very recently: specifically, the ideal-glass-like $\phi_{\rm J} \simeq 0.909$ packings examined in Ref.\ \cite{fan26} had $\lambda \simeq 2.36$.
Here we found $\lambda_{\rm xtal}(\alpha) < 2$ for all $1.025 \lesssim \alpha \lesssim 1.42$, and $\lambda_{\rm xtal}(\alpha) < 2.36$ for all $\alpha \lesssim 1.98$.

We also identified some of the key mechanisms via which polydisperse ellipses are able to form disordered $\phi > \phi_{\rm xtal}^{\rm mono}$ packings.
For example, the results presented in Section \ref{subsec:IIIE} suggest the following picture for low-dispersity ellipses with $\alpha \simeq \alpha_{\rm max}$:
The majority of the particles group themselves into moderately-fractionated, plastic-crystalline grains.
Since these grains (by definition) lack nematic order, they tend to be irregularly shaped.
Thus they are geometrically frustrated and cannot tile space.
However, the remaining particles (primarily the smallest and largest particles) are able to arrange themselves in such a way as to optimally pack the regions \textit{between these grains}.
Critically, the use of SWAP moves during slow particle growth appears to optimize \textit{both} of these packing processes.
Comparable structure was reported in the $\alpha \simeq \alpha_{\rm max}$ packings discussed in Ref.\ \cite{hoy24b}, but the combination of larger $\lambda$ values and slower particle growth employed here allowed these processes to produce denser-than-crystalline packings.

\begin{figure}[h]
\centering
\includegraphics[width=3.375in]{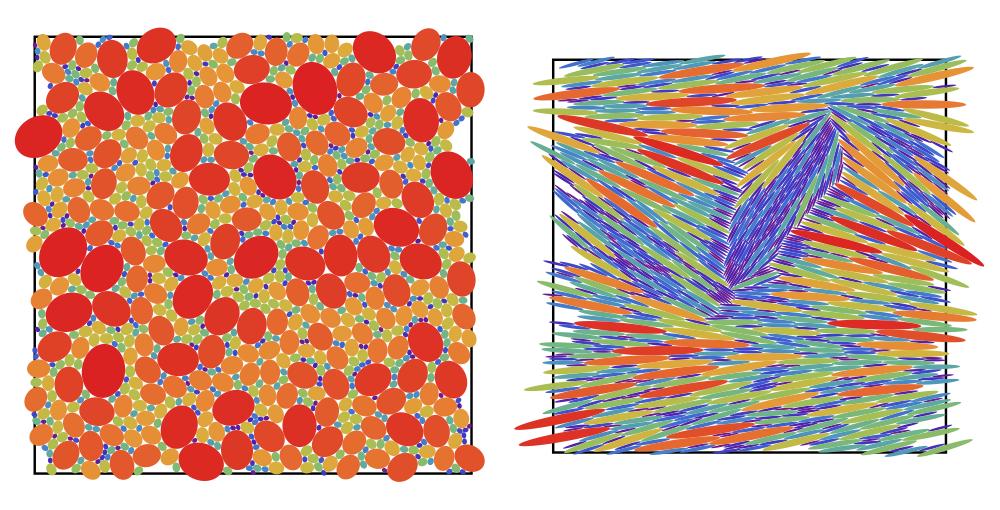}
\caption{Examples of $\phi > \phi_{\rm xtal}$ packings with $N = 10^3$ but larger $\alpha$ or $\lambda$ values than were discussed above.  The left and right panels respectively have $(\alpha,\lambda) = (5/4,10)$ and $(10, 5)$, $\phi_{\rm J} = 0.9310$ and $0.9095$, and $Z_{\rm J} = 5.928$ and $5.824$.}
\label{fig:Z}
\end{figure}

Finally, we also examined several aspects of these packings which are independent of whether they are denser-than-crystalline, but should be useful in the intense ongoing effort to develop ultrastable amorphous solids.\cite{bolton26,fan26,hoy25,berthier24,berthier26,galliano26,dyre26}
For example, we anticipate that the differences between the mechanical properties of the ultradense ellipse packings discussed above and those of ``normal'' ellipse packings generated using fast particle growth\cite{donev07,delaney05,rocks23,jiang26,vanderwerf18,mailman09,schreck10,schreck12} will be comparable to the differences between ultrastable and ``normal'' disk and sphere packings.
More specifically, we believe that (when converted into soft-particle packings) they will have much higher bulk and shear moduli as well as higher melting temperatures than their normal counterparts.
Those that are also triangulated and effectively-hyperuniform may also have Debye-like $D(\omega)$ in the $\omega \to 0$ limit.\cite{bolton26,fan26}

\begin{figure*}
\centering
\includegraphics[width=7in]{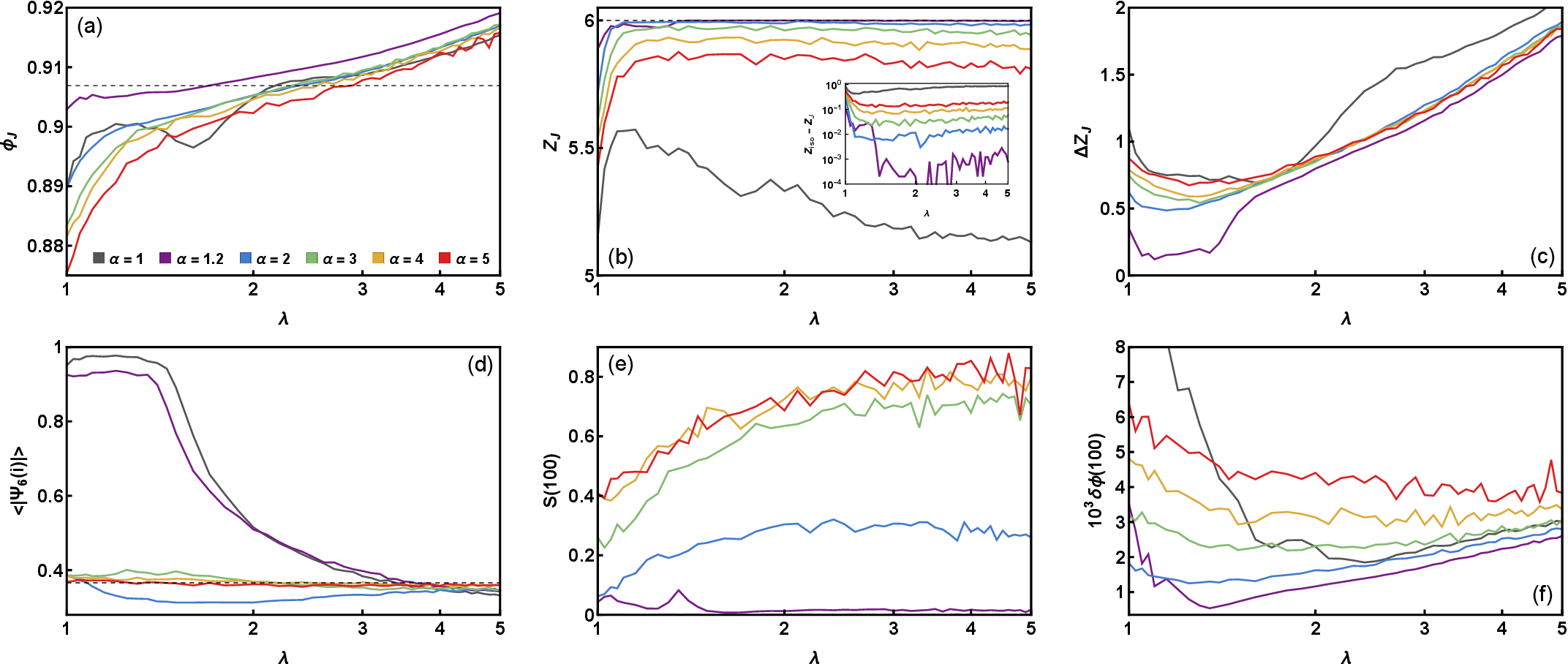}
\caption{Variation of jamming densities [panel (a)], coordination numbers  [panel (b)], standard deviations of the coordination numbers [panel (c)],  average single-particle hexatic order [panel (d)], average intermediate-range nematic order [panel (e)], and average intermediate-range packing-fraction fluctuations [panel (f)] with $\lambda$, for selected $\alpha$.
Dashed lines in panels (a), (b) and (d) respectively indicate $\phi = \phi_{\rm xtal}$, $Z_{\rm J} = Z_{\rm iso}$ and $\langle|\Psi_6|\rangle_{\rm rand}$.}
\label{fig:appfig}
\end{figure*}

Intensive followup study will be required to determine whether these hypotheses are indeed correct, and many related questions about the nature of ultrastable anisotropic-particle packings remain open.
For example, one crucial difference between the triangulated ellipse packings reported above and those discussed in Ref.\ \cite{bolton26} is that triangulated ellipse packings are exactly-isostatic whereas triangulated disk packings are strongly hyperstatic.
Will this difference alter the effects of triangulation on system's mechanical/vibrational/thermal properties, and if so, how?
We plan to explore such questions in upcoming work.

We conclude by emphasizing that while we have focused in this paper on systems with $\alpha \leq 5$ and $\lambda \leq 5$, our algorithm is also capable of generating $\phi > \phi_{\rm xtal}$ packings with larger $\alpha$ and/or $\lambda$.
Figure \ref{fig:Z} shows two examples.
Systems with small $\alpha$ and large $\lambda$ appear very disordered, while the high-$\alpha$ limit yields fractionated nematic polycrystals.

\section{Appendix: Variation of structure with $\lambda$ for selected $\alpha$}
\label{sec:appA}

Figure \ref{fig:appfig} shows how the six key structural metrics examined in Sections \ref{subsec:IIIA}-\ref{subsec:IIIB} (Figs.\ \ref{fig:phiandZ1} and \ref{fig:spatiallyaveragedmetrics}) vary with $\lambda$ for six representative values of $\alpha$.

Panel (a) shows that (with the exception of disks, and modulo statistical noise) all systems' $\phi_{\rm J}$ increase monotonically with $\lambda$.
Disks' $\phi_{\rm J}(\lambda)$ exhibit a maximum at $\lambda \simeq 1.25$ and a minimum at $\lambda \simeq 1.6$.
The former corresponds to a maximum in SWAP's ability to generate fractionated crystals, while the latter corresponds to the point at which the diminishing strength of this effect is balanced by the increase of \textit{disordered} packings' $\phi_{\rm J}(\lambda)$.

Panel (b) shows that our algorithm generates broad plateaus in $Z_{\rm J}(\alpha,\lambda)$ for all $\alpha \gtrsim \alpha_{\rm max}$, but that the height of these plateaus is strongly $\alpha$-dependent.
The negative $\partial Z_{\rm J}/\partial \alpha$ for $\alpha \gtrsim \alpha_{\rm max}$ may result from limitations of the algorithm; future work will attempt to identify and eliminate these.

Panel (c) shows that the shapes of the $\Delta Z_{\rm J}(\alpha,\lambda)$ curves depend strongly on $\alpha$ only for $\alpha < 2$.
The global minimum in $\Delta Z_{\rm J}(\alpha,\lambda)$ is $\Delta Z_{\rm J}(1.2,1.1) \simeq 0.2$.
Intriguingly, minimizing $\Delta Z_{\rm J}$ does \textit{not} always correspond to maximizing $Z_{\rm J}$; $Z_{\rm J}(1.2,1.1) = 5.986$.

Panel (d) shows that for $\alpha \lesssim \alpha_{\rm max}$, increasing $\lambda$ reduces local hexatic order only for $\lambda > 1.4$.
The maxima in  $\langle|\Psi_6(i)|\rangle$ occur at $\lambda = 1.2$ for both $\alpha = 1$ and $\alpha = 1.2$.
This shows that SWAP's ability to generate fractionated \textit{hexatic} crystals for small $\alpha$ is independent of whether these crystals are also \textit{hexagonal}.
For $\alpha \gtrsim 2$, hexatic order remains minimal for all $\lambda$.

Panel (e) shows that (modulo statistical noise) all systems' $S_{\rm loc}(100)$ increase monotonically with $\lambda$.
This result is counterintuitive, as larger $\lambda$ might be expected to uppress the formation of intermediate-range nematic order.
Instead it appears that larger $\lambda$ enhance SWAP's ability to generate efficiently-packed locally-nematic domains.

Finally, panel (f) shows that the $\alpha$- and $\lambda$-dependencies of intermediate-range packing-fraction fluctuations are strongly coupled.
For $\alpha \lesssim \alpha_{\rm max}$, the curves exhibit strong minima at $\lambda$ values that decrease with increasing $\alpha$.
As $\alpha$ increases further, these minima first weaken and then (modulo statistical noise) disappear entirely for $\alpha \gtrsim 3$.
This indicates that the connection between maximization of $\phi_{\rm J}$ and minimization of $\delta \phi(\mathcal{N})$ (Eq.\ \ref{eq:phiflucts1}) is only strong in systems that do not form strong intermediate-range nematic order.

Notably, \textit{none} of these metrics exhibit rapid changes as systems cross the $\lambda = \lambda_{\rm xtal}$ boundary shown in Fig.\ \ref{fig:xtal}.
Combined with the results shown in Section \ref{sec:results}, this data reinforces our claim that the formation of packings with $\phi > \phi_{\rm xtal}^{\rm mono}$ does not require any easily-definable structural transition.

\section*{Author contributions}

Kinsley Trlak: Investigation, Formal analysis,  Software, Visualization.
Dewni de Silva: Investigation, Formal analysis.
Salvatore Torquato: Conceptualization, Methodology, Writing -- review \& editing.
Robert Hoy: Conceptualization, Methodology, Software, Investigation, Formal analysis, Visualization, Writing -- original draft, Writing -- review \& editing, Supervision.


\section*{Conflicts of interest}

There are no conflicts to declare.

\section*{Data availability}

The processed data appearing in this article, the particle coordinates for the packings discussed above, and the \texttt{C++} code used to generate these packings are available at \underline{http://TBA}.


\section*{Acknowledgements}

This material is based upon work supported by the National Science Foundation under Grant No.\ DMR-2026271.
S.T.\ was supported by the U.S. Army Research Office under Cooperative Agreement No.\ W911NF-22-2-0103.



\balance


\bibliographystyle{rsc}

\providecommand*{\mcitethebibliography}{\thebibliography}
\csname @ifundefined\endcsname{endmcitethebibliography}
{\let\endmcitethebibliography\endthebibliography}{}

\end{document}